\documentclass[aps,pra,amsmath,amssymb,graphicx,preprint,citeautoscript]{revtex4-1}
\usepackage{color}
\usepackage{verbatim}
\usepackage{mciteplus}
\usepackage{amsbsy}
\usepackage{amstext}
\usepackage{amssymb}
\usepackage{float}
\usepackage{geometry}
\usepackage{setspace}
\usepackage{xkeyval}
\usepackage{graphicx}
\usepackage[caption=false]{subfig}
\usepackage{placeins}
\usepackage{colortbl}
\usepackage{gensymb}

\usepackage[unicode=true, bookmarks=true, bookmarksnumbered=false, bookmarksopen=false,
  breaklinks=false,pdfborder={0 0 0},backref=false,colorlinks=true]{hyperref}

\hypersetup{pdftitle={First principles electron-correlated calculations of optical absorption in magnesium clusters},
pdfauthor={Ravindra Shinde, Alok Shukla}, pdfsubject={Computational Material Science},
unicode=false,pdftoolbar=true,pdfmenubar=true,pdffitwindow=true,pdfstartview={FitH},pdfnewwindow=true,pdfcreator={RevTeX},linkcolor=red,citecolor=blue,urlcolor=blue}
  
\begin{document}


\title{First principles electron-correlated calculations of optical absorption in magnesium clusters} 

\author{Ravindra Shinde}
\email{ravindra@mrc.iisc.ernet.in}
\affiliation{Materials Research Center, Indian Institute of Science, Bangalore 560012, India.}

\author{Alok Shukla}
\email{shukla@phy.iitb.ac.in}
\affiliation{Department of Physics, Indian Institute of Technology Bombay, Mumbai  400076, India.}

\date{\today}

\begin{abstract}
In this paper, we report large-scale configuration interaction (CI) calculations of linear optical absorption spectra of various isomers of magnesium clusters Mg$_{n}$ (n=2--5), corresponding to valence transitions. Geometry optimization of several low-lying isomers of each cluster was carried out using coupled-cluster singles doubles (CCSD) approach, and these geometries were subsequently employed to perform ground and excited state calculations using either the full-CI (FCI) or the multi-reference singles-doubles configuration interaction (MRSDCI) approach, within the frozen-core approximation. Our calculated photoabsorption spectrum of magnesium dimer (Mg$_{2}$) isomer is in excellent agreement with the experiments both for peak positions, and intensities. Owing to the sufficiently inclusive electron-correlation effects, these results can serve as benchmarks against which future experiments, as well as calculations performed using other theoretical approaches, can be tested. 
\end{abstract}


\maketitle

\section{\label{sec:introduction}INTRODUCTION}

Clusters of group II elements, such as magnesium, are particularly interesting because they have two valence electrons, quasi-filled closed shells,
and in bulk they are metals. In the case of small clusters, the bonding between atoms is expected to be of van der Walls type. This is evident
in the case of extensively studied magnesium dimer. It exhibits a large bond length of 3.92 \AA{} and 0.034 eV/atom binding energy. However,
it is seen that for larger clusters this bonding becomes stronger. Thus, the study of divalent metals is appropriate for the evolution of various
cluster properties and to test various theoretical methods. Involvement of metal atoms in the clusters makes theoretical treatment a demanding
task, mainly because of several nearly degenerate electronic states. In such situations, only multi-reference configuration interaction
methods or coupled cluster singles doubles with perturbative triples (CCSD(T)) is known to provide best qualitative results \cite{ahlrichs_mg_pccp}.
Since in this paper, we are dealing with small-sized clusters of magnesium, treated at a large-scale multi-reference configuration interaction
singles doubles level of theory, the results will be superior to other \emph{ab initio} quantum chemical methods.

There have been a large number of studies of equilibrium geometries and electronic structure of small magnesium clusters \cite{ahlrichs_mg_pccp,ele-struct-magnesium-pra,car_kumar_magnesium_prb,exp_mg_dimer_spectra_jcp,ground_excited_mg2_jcp,jellinek_mg2-mg5_jpca,manninen_mg_evolution_epjd,kaplan_mg3_jcp}. Andrey \emph{et al.}\cite{ele-struct-magnesium-pra} studied the evolution of the electronic structure of magnesium clusters with cluster size using all-electron density functional theoretical method. An evolution from non-metal to metal was explained using a gradient-corrected DFT calculations by Jellinek and Acioli \cite{jellinek_mg2-mg5_jpca}, and by Akola \emph{et al} \cite{manninen_mg_evolution_epjd}. Larger clusters were studied at DFT level by Kohn \emph{et al.}\cite{ahlrichs_mg_pccp}
Kumar and Car performed \emph{ab initio} density functional molecular dynamics study of smaller magnesium clusters within local density
approximation \cite{car_kumar_magnesium_prb}. Stevens and Krauss calculated electronic structure of ground and excited states of Mg dimer using
multiconfigurational self-consistent field method \cite{ground_excited_mg2_jcp}. Kaplan, Roszak, and Leszczynski investigated the nature of binding
in the magnesium trimer at MP4 level \cite{kaplan_mg3_jcp}. 

The optical absorption in dimer was studied experimentally by McCaffrey and Ozin \cite{exp_mg_dimer_spectra_jcp}, and  Lauterwald and Rademann \cite{mgdimerexp2} in Ar, Kr and Xe matrices,  while Balfour and Douglas \cite{exp_mg_dimer_douglas} measured it in the gas phase. Solov'yov \emph{et al.} calculated optical absorption spectra of global minimum structures of magnesium clusters using TDDFT and compared the spectra with results of classical Mie theory \cite{optical_mg_jpb}. However, to best of our knowledge, no other experimental or theoretical study exists for optical absorption and excited states calculations of various low-lying isomers of magnesium clusters. The distinction of different isomers of a cluster has to be made using experimental or theoretical techniques based upon properties which are shape and size dependent, unlike, mass spectroscopy which depends only on the mass of the cluster.  We have addressed this issue by performing large-scale correlated calculations of optical absorption spectra of various isomers
of magnesium clusters Mg$_{n}$ (n=2--5), at MRSDCI level of theory.  Hence, our theoretical results can help in distinguishing between different isomers of a cluster, when coupled with the experimental measurements of their optical absorption. We also investigate the nature of optical
excitations by analyzing the wavefunctions of various excited states. Furthermore, wherever possible, the results have been compared with the available literature. In earlier works, we reported similar calculations of  optical absorption spectra of various isomers of small boron and aluminum clusters \cite{smallboron,aluminum-ravi}.

\section{\label{sec:theory}THEORETICAL AND COMPUTATIONAL DETAILS}

\begin{figure*}
\centering
\subfloat[\textbf{Mg$_{\mathbf{2}}$, D$_{\mathbf{\boldsymbol{\infty}h}}$,
$^{\mathbf{1}}\mathbf{\boldsymbol{\Sigma_{g}}}$ \label{subfig:subfig-mg2}}]{\includegraphics[width=2.5cm]{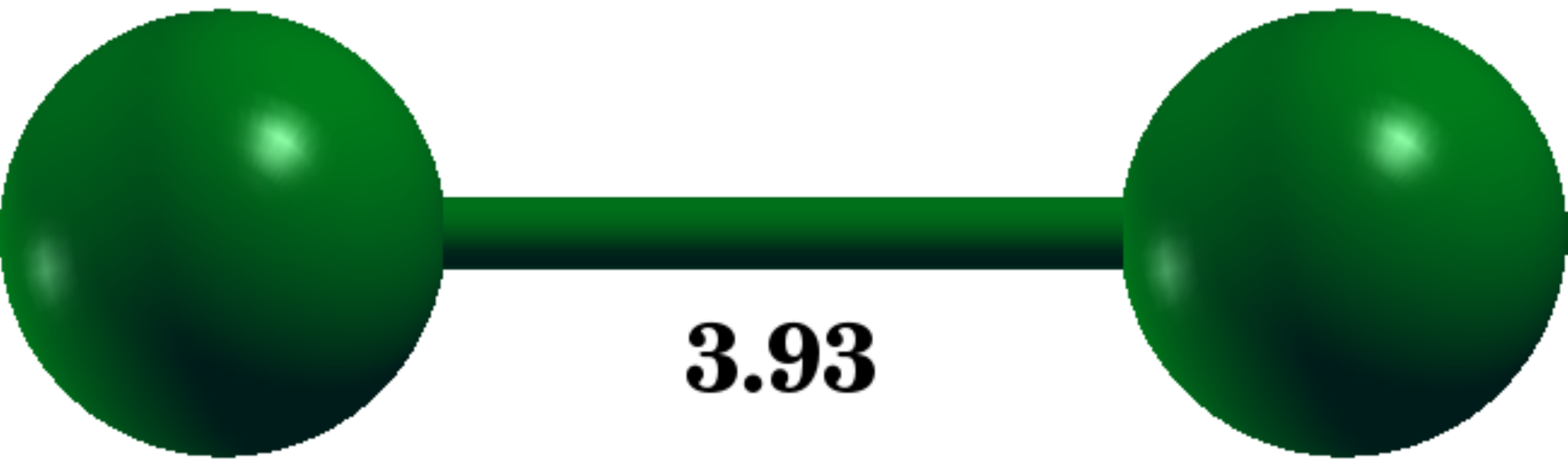}
}\hfill\subfloat[\textbf{Mg$_{\boldsymbol{3}}$, D$_{\boldsymbol{3h}}$, $\boldsymbol{^{1}A_{1}^{'}}$\label{subfig:subfig-mg3-equil}}]{\includegraphics[width=2.3cm]{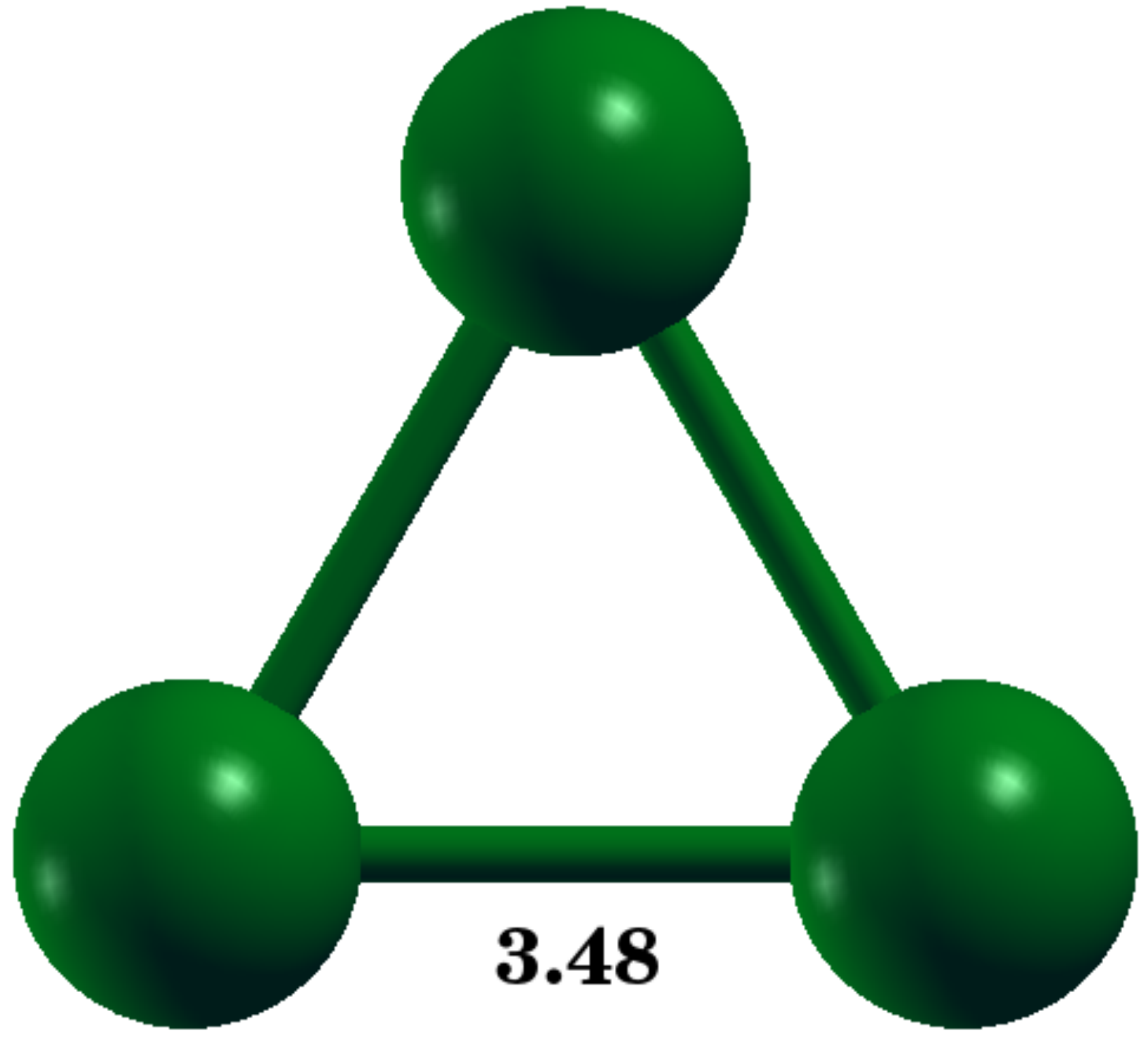}
}\hfill\subfloat[\textbf{Mg$_{\boldsymbol{3}}$, D$_{\mathbf{\boldsymbol{\infty}h}}$,
$\boldsymbol{^{3}\Pi_{u}}$\label{subfig:subfig-mg3-linear}}]{\includegraphics[width=3.4cm]{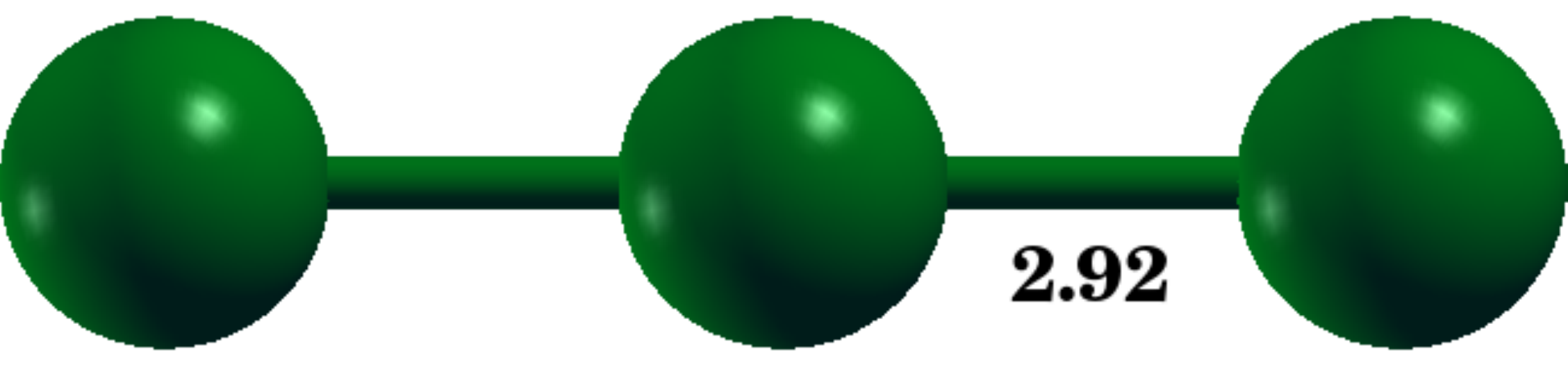}
}\hfill\subfloat[\textbf{Mg$_{\boldsymbol{3}}$, C}$_{\boldsymbol{2v}}$, $\boldsymbol{^{3}A_{2}}$\label{subfig:subfig-mg3-iso1}]{\includegraphics[width=2.0cm]{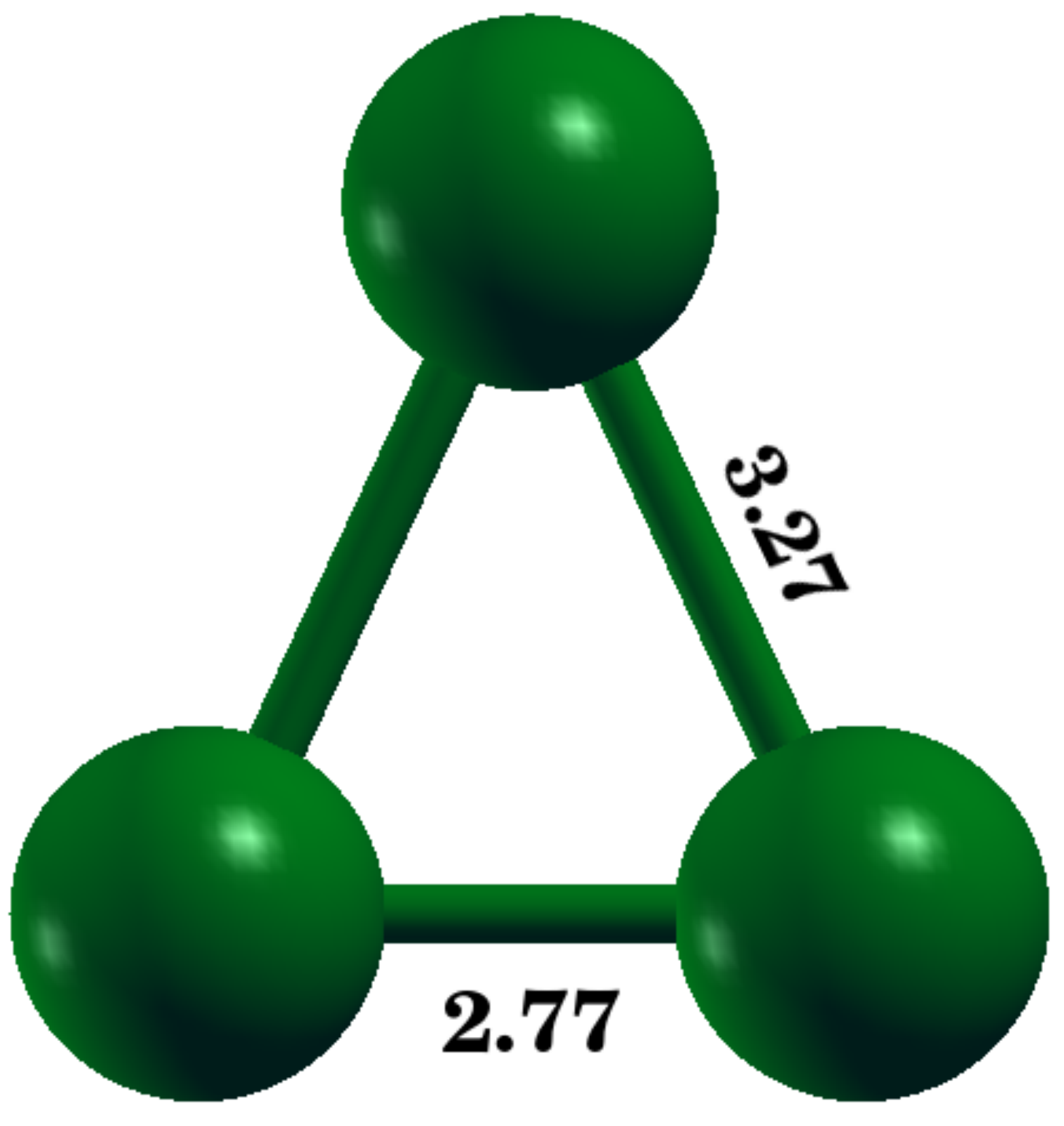}
}\\
\subfloat[\textbf{Mg$_{\boldsymbol{3}}$, C$_{\boldsymbol{2v}}$, $\boldsymbol{^{3}B_{1}}$\label{subfig:subfig-mg3-iso2}}]{\includegraphics[width=2.5cm]{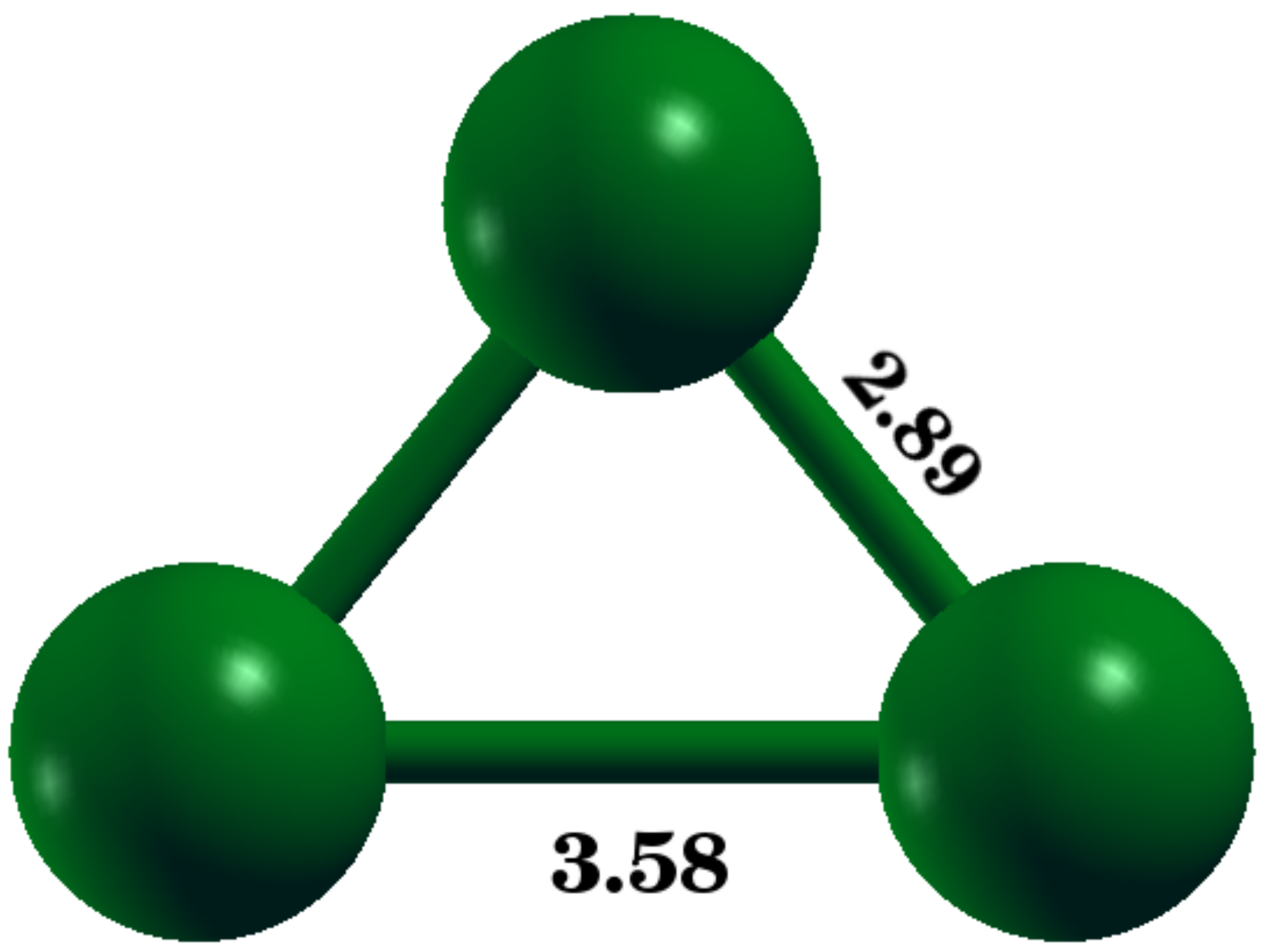}
}\hfill\subfloat[\textbf{Mg$_{\boldsymbol{4}}$, T$_{\boldsymbol{d}}$, $\boldsymbol{^{1}A{}_{1}}$\label{subfig:subfig-mg4-pyramidal}}]{\includegraphics[width=2.3cm]{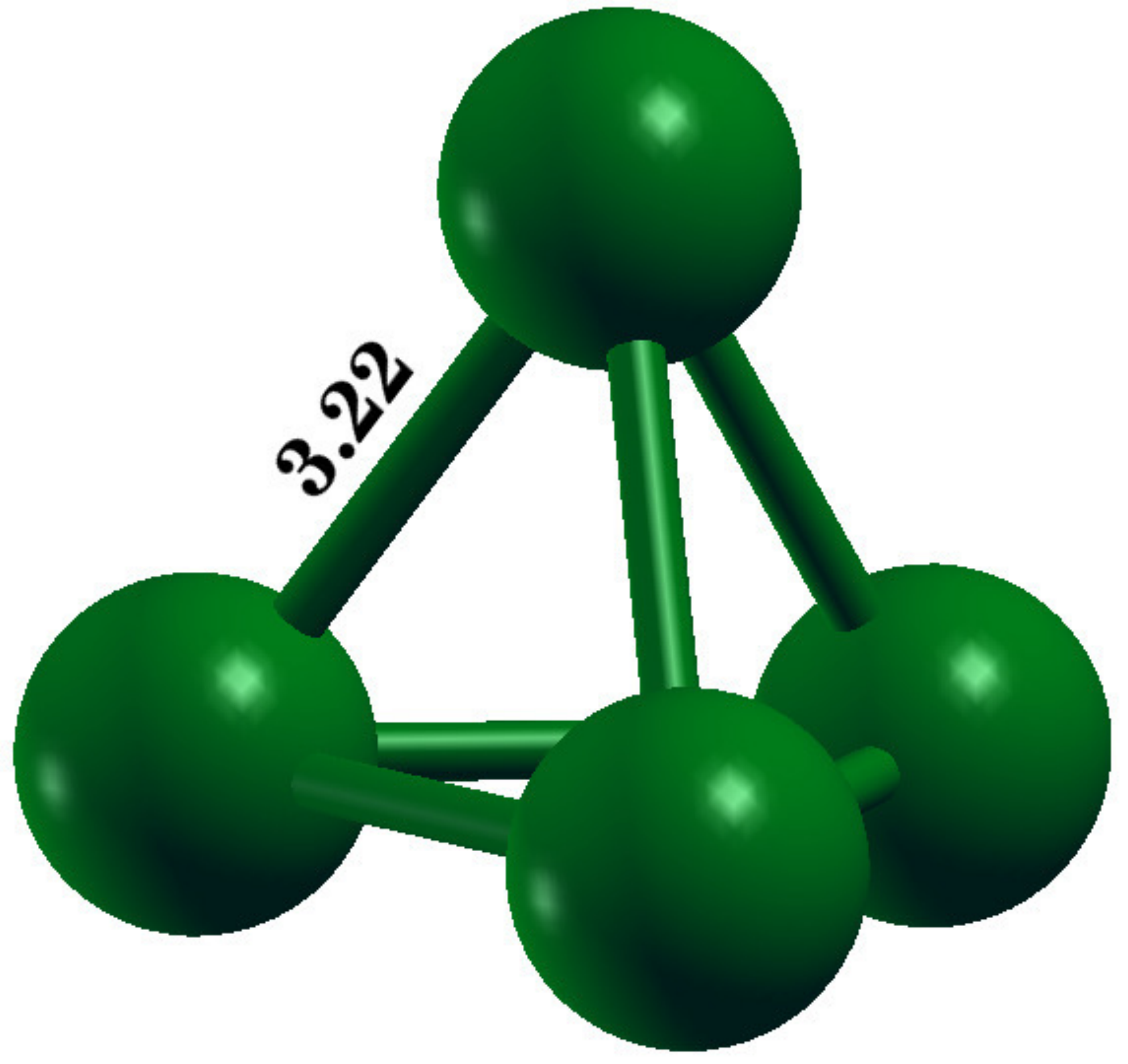}
}\hfill\subfloat[\textbf{Mg$_{\boldsymbol{4}}$, D$_{\boldsymbol{2h}}$, $\boldsymbol{^{3}B{}_{3u}}$\label{subfig:subfig-mg4-rhombus}}]{\includegraphics[width=3.2cm]{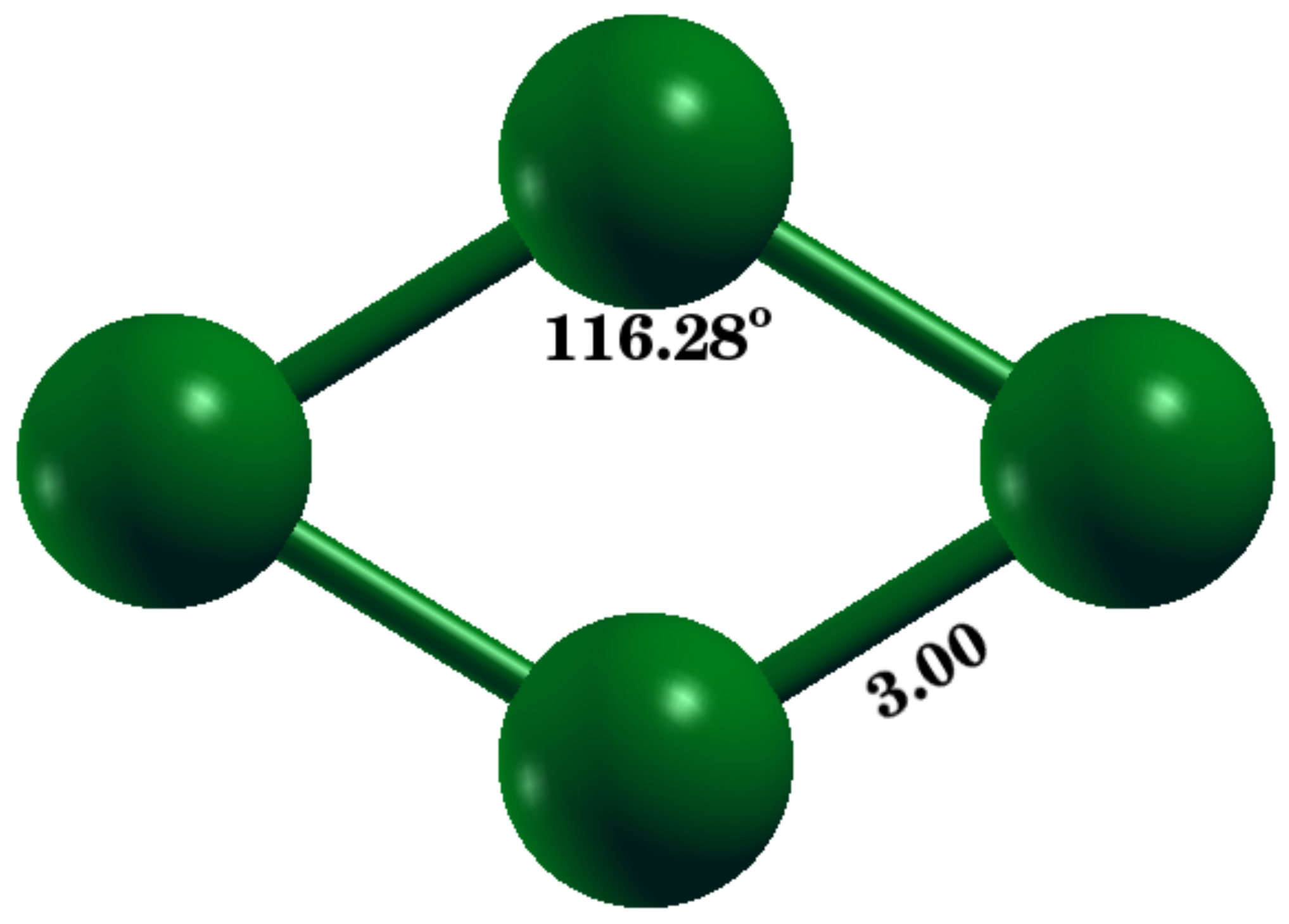}
}\hfill\subfloat[\textbf{Mg$_{\boldsymbol{4}}$, D$_{\boldsymbol{4h}}$,} $\boldsymbol{^{3}A{}_{g}}$\label{subfig:subfig-mg4-square}]{\includegraphics[width=2.1cm]{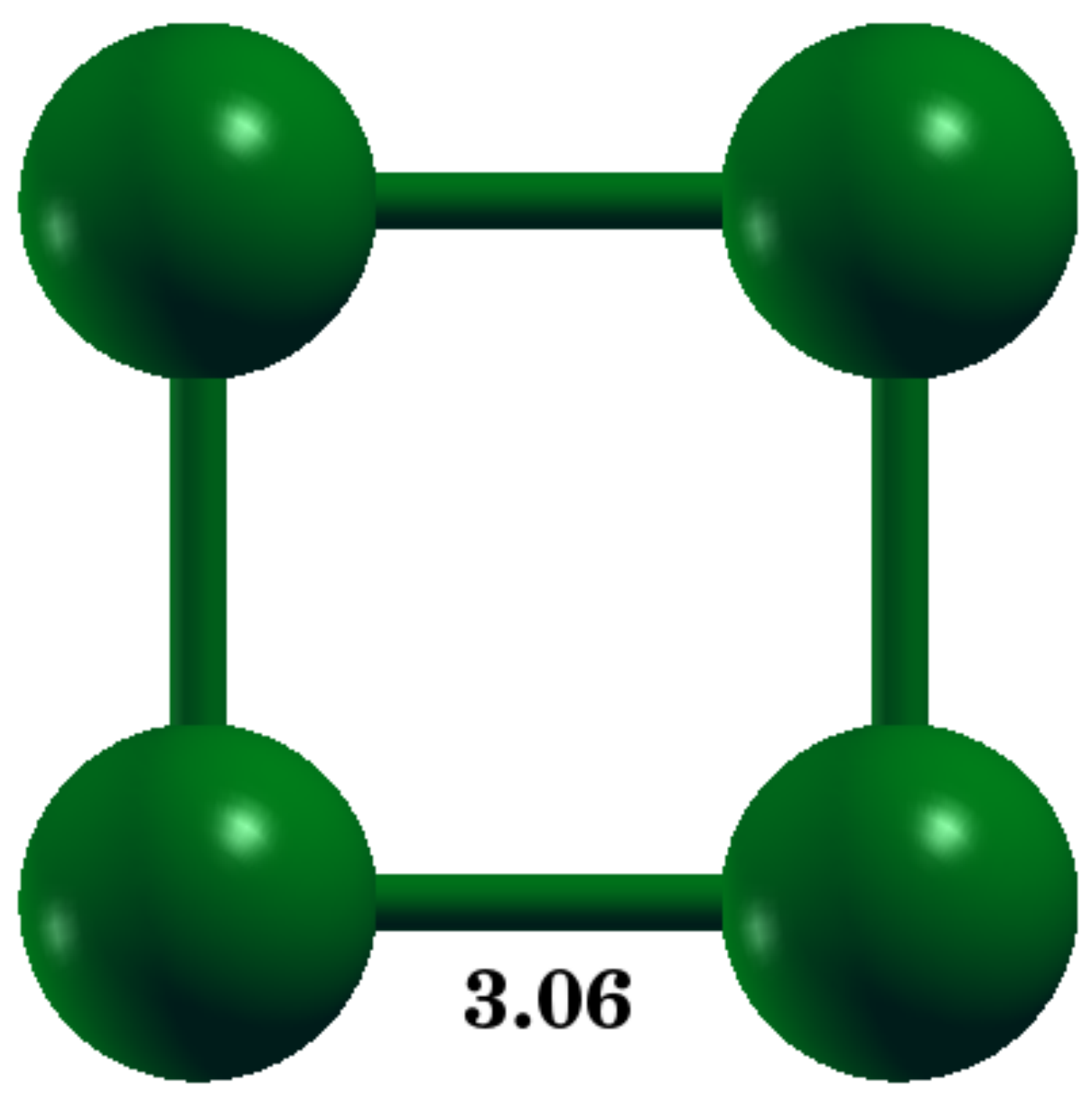}
}\\
\subfloat[\textbf{Mg$_{\boldsymbol{5}}$, D$_{\boldsymbol{3h}}$, $\boldsymbol{^{1}A{}_{1}^{'}}$\label{subfig:subfig-mg5-bipyramid}}]{\includegraphics[width=2.4cm,angle=90]{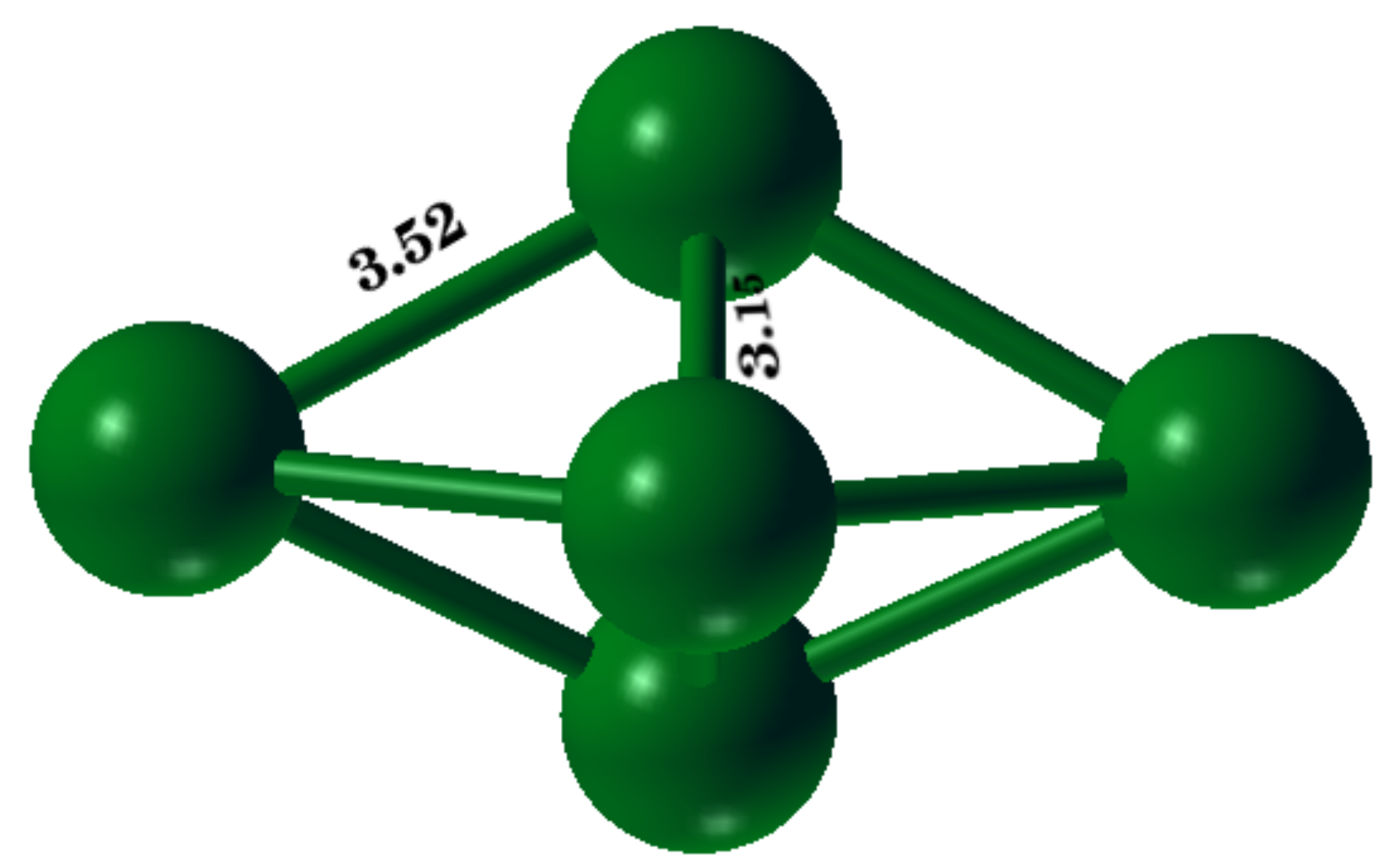}
}\hfill\subfloat[\textbf{Mg$_{\boldsymbol{5}}$, C$_{\boldsymbol{4v}}$,} $\boldsymbol{^{1}A{}_{1}}$\label{subfig:subfig-mg5-pyramid}]{\includegraphics[width=3.5cm]{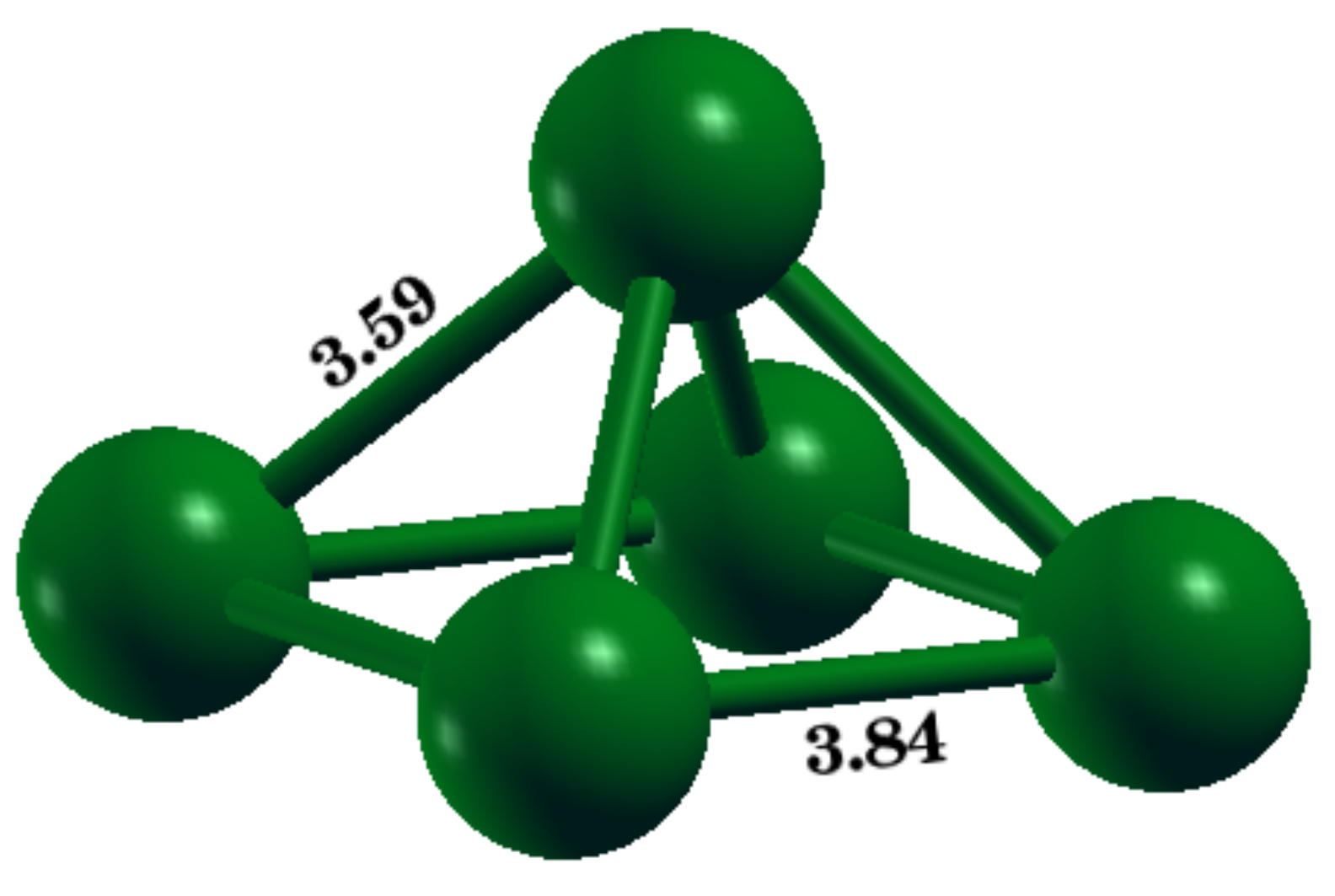}
}
\vspace{0.1cm}
 \caption{\label{fig:geometry-magnesium} Optimized geometries of Mg clusters considered in this work, along with the symmetries of their ground state wave functions. Geometry optimization was carried out at the CCSD level, and all lengths are in \AA{} units. }
\end{figure*}

A size-consistent coupled-cluster singles doubles (CCSD) level of theory along with a 6-311+G(d) basis set was used for geometry optimization, followed by vibrational analysis \cite{gaussian09}. This basis set is well-suited for the ground state calculations. Different spin multiplicities of the isomers were taken into account for the optimization to determine the true ground state geometry. The process of optimization was initiated by using the geometries reported by Lyalin \emph{et al.}\cite{ele-struct-magnesium-pra}, based upon first principles DFT-based calculations. The final optimized geometries of the isomers are shown in Fig. \ref{fig:geometry-magnesium}.

For computing the optical absorption spectra, both ground and excited state wave functions for these optimized geometries were calculated
using multireference singles-doubles configuration interaction (MRSDCI) method \cite{meld}. This approach consists of generating singly-
and doubly-substituted configurations from a set of reference configurations, which are chosen based upon their contribution to the targeted wave
functions obtained from a lower-level calculation based upon, say, single-reference singles-doubles configuration interaction (SDCI)
method. Optical absorption spectra are computed at each stage of the calculation, and the targeted wave functions are analyzed to examine
whether more reference configurations are needed. This procedure is repeated until the absorption spectrum of the system under consideration
converges. Such an approach is equally efficient both for ground and excited state calculations because it takes into account the electron
correlation effects for all the targeted states in an individualized manner, something which is not possible in single reference approaches.  The transition dipole moment matrix elements are calculated using these ground- and excited-state wavefunctions, and are subsequently utilized to compute linear optical absorption spectrum assuming a Lorentzian line shape. The numerical approach described here has been extensively used in our earlier works dealing
with the optical properties of conjugated polymers \cite{mrsd_prb_02,mrsd_prb_05,mrsd_prb_07,mrsd-himanshu-jpca13,mrsd-himanshu-jcp14},
as well as atomic clusters \cite{smallboron,aluminum-ravi}. For the smallest cluster, namely, Mg dimer, it was possible to use the full CI approach, within the frozen-core approximation.

Since the computational effort involved in a CI calculation scales $\approx N^{6}$, where $N$ is the total number of orbitals
involved in the calculation, it can become intractable if a large basis set, leading to a large number of molecular
orbitals (MOs) is employed. To reduce the MO basis set size, we employed the so-called ``frozen-core approximation'', in which no virtual
transitions are allowed from the chemical core of magnesium atoms, thereby leading to two valence electrons per atom, which were treated
as active during the calculations. Furthermore, an upper limit of one hartree on the energies of the virtual orbitals to be included in the calculations
was imposed, so as to control the size of the CI expansion without compromising the accuracy of the optical absorption spectrum. In the next section, we carefully examine the effects of all these approximations on our calculations. 

Further computational efficiency was achieved by making full use of point-group symmetries  (D$_{2h}$, and its subgroups), wherever applicable.

\subsection{Choice of Basis Set}

\begin{figure}
\centering \includegraphics[width=1\columnwidth]{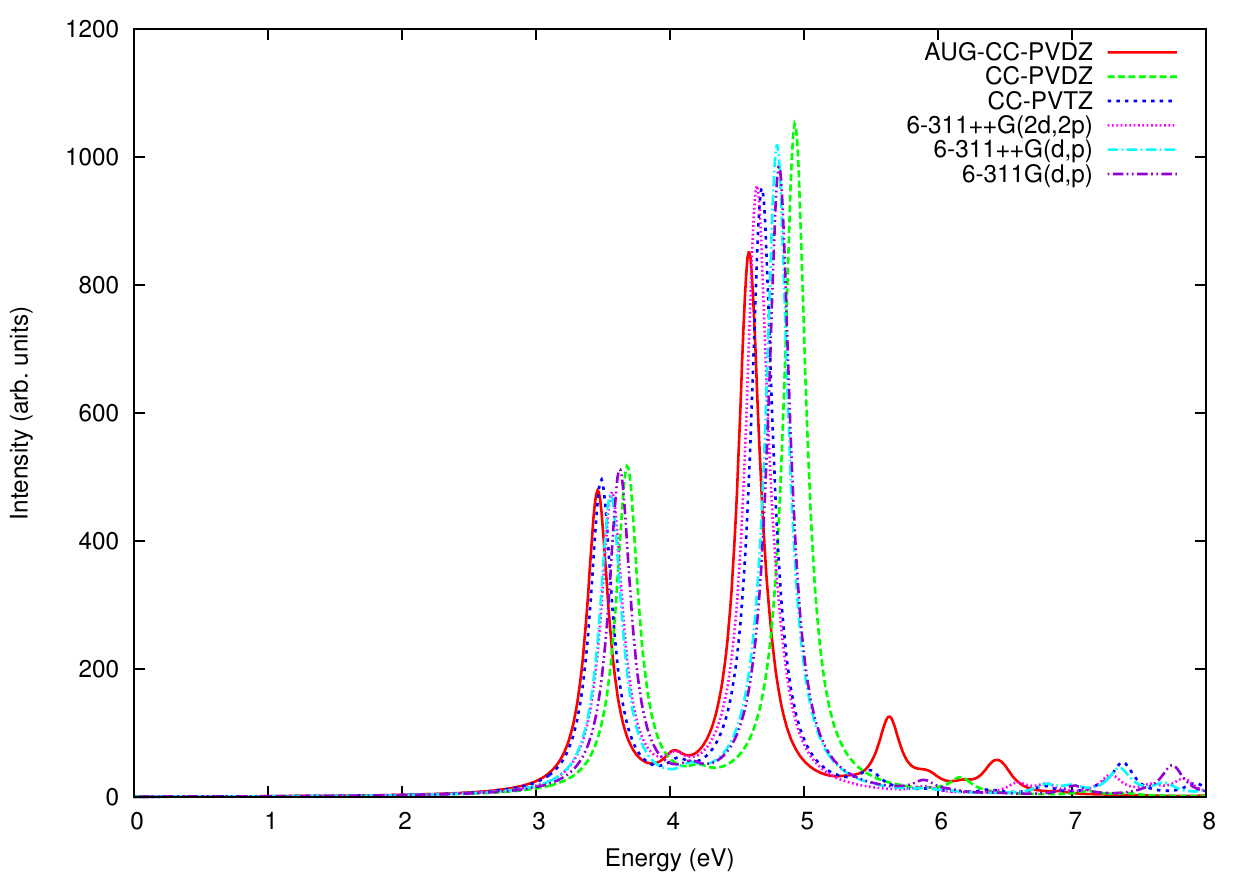}
\vspace{0.2cm}
 \caption{\label{fig:magnesium-basis-study}Optical absorption in Mg$_{2}$
calculated using various Gaussian contracted basis sets.}
\end{figure}

Electronic structure calculations generally depend upon the size and the quality of basis set used. To explore the basis set dependence
of computed spectra, we used several basis sets \cite{emsl_bas1,emsl_bas2,emsl-mg-basis} to compute the optical absorption spectrum of the magnesium dimer.
For the purpose, we used basis sets named aug-cc-pVDZ, cc-pVDZ, cc-pVTZ, 6-311++G(2d,2p), 6-311++G(d,p) and 6-311G(d,p), which consist of polarization
functions along with diffuse exponents \cite{emsl_bas1,emsl_bas2,emsl-mg-basis}. From the calculated spectra presented in Fig. \ref{fig:magnesium-basis-study}
the following trends emerge: the spectra computed by various correlation consistent basis sets (aug-cc-pVDZ, cc-pVDZ, cc-pVTZ) are in good
agreement with each other in the energy range up to 5 eV, while those obtained using the other basis sets (6-311++G(2d,2p), 6-311++G(d,p)
and 6-311G(d,p)) disagree with them substantially, particularly in the higher energy range. Peaks at 5.6 eV and 6.5 eV are seen only
in the spectrum calculated using augmented basis set. Because of the fact that augmented basis sets are considered superior for molecular
calculations, we decided to perform calculations on the all the clusters using the aug-cc-pVDZ basis set.

\subsection{Size of the CI Expansion}

The electron correlation effects, both in ground state as well as excited states, were taken into account in our calculations by the inclusion
of relevant configurations in the reference space of MRSDCI expansion. Larger the reference configuration space, larger will be the CI expansion,
which is prohibitive for bigger systems. A good chemical accuracy can be achieved by moderately sized CI expansions within the MRSDCI
approach, provided the reference configurations are chosen carefully. In Table \ref{tab:energies-irrep-magnesium} we present the average
number of reference states (N$_{ref}$) included in the MRSDCI expansion and the average number of configurations (N$_{total}$) for different
isomers.  The average is computed over different irreducible representations required in the calculation of the ground and various excited states of a given isomer. Large scale nature of these calculations is obvious from the fact that the total number of configurations in the CI expansion, N$_{total}$, ranges
from $\approx$ 45000 for the smallest cluster (Mg$_{2}$), to around three million for each symmetry subspace of Mg$_{5}$, implying that
the electron-correlation effects have been adequately included.

Before we discuss the absorption spectrum for each isomer, we present
the ground state energies along with the relative energies of each
isomer are given in Table \ref{tab:energies-irrep-magnesium}.

\begin{table*}
\caption{The average number of total configurations (N$_{total}$) involved
in MRSDCI calculations, ground state (GS) energies (in Hartree) at
the MRSDCI level and relative energies (in eV) of various isomers of magnesium clusters.\label{tab:energies-irrep-magnesium}}
\begin{ruledtabular}
\begin{tabular}{clcccc}
Cluster  & Isomer  & N$_{ref}$  & N$_{total}$  & GS energy  & Relative \tabularnewline
 &  &  &  & (Ha)  & energy (eV) \tabularnewline
\hline 
 &  &  &  &  & \tabularnewline
Mg$_{2}$  & (Fig. \ref{fig:geometry-magnesium}\subref{subfig:subfig-mg2}) Linear  & 1\footnotemark[1]  & 44796  & -399.2847413  & 0.00 \tabularnewline
 &  &  &  &  & \tabularnewline
Mg$_{3}$  & (Fig. \ref{fig:geometry-magnesium}\subref{subfig:subfig-mg3-equil}) Equilateral Triangular  & 30  & 239465  & -598.9270344  & 0.00 \tabularnewline
 & (Fig. \ref{fig:geometry-magnesium}\subref{subfig:subfig-mg3-linear}) Linear  & 55  & 460187  & -598.8759291  & 1.39 \tabularnewline
 & (Fig. \ref{fig:geometry-magnesium}\subref{subfig:subfig-mg3-iso1}) Isosceles Triangular-1  & 34  & 516337  & -598.8569875  & 1.91\tabularnewline
 & (Fig. \ref{fig:geometry-magnesium}\subref{subfig:subfig-mg3-iso2}) Isosceles Triangular-2  & 32  & 359780  & -598.8093768  & 3.20\tabularnewline
 &  &  &  &  & \tabularnewline
Mg$_{4}$  & (Fig. \ref{fig:geometry-magnesium}\subref{subfig:subfig-mg4-pyramidal}) Pyramidal  & 32  & 2962035  & -798.5781385  & 0.00 \tabularnewline
 & (Fig. \ref{fig:geometry-magnesium}\subref{subfig:subfig-mg4-rhombus}) Rhombus  & 29  & 1278632  & -798.5405148  & 1.02 \tabularnewline
 & (Fig. \ref{fig:geometry-magnesium}\subref{subfig:subfig-mg4-square}) Square  & 35  & 1319301  & -798.5278160  & 1.37 \tabularnewline
 &  &  &  &  & \tabularnewline
Mg$_{5}$  & (Fig. \ref{fig:geometry-magnesium}\subref{subfig:subfig-mg5-bipyramid}) Bipyramidal  & 11  & 3242198  & -998.2044402  & 0.00 \tabularnewline
 & (Fig. \ref{fig:geometry-magnesium}\subref{subfig:subfig-mg5-pyramid}) Pyramidal  & 28  & 2215749  & -998.1980062  & 0.18 \tabularnewline
 &  &  &  &  & \tabularnewline
\end{tabular} 
\end{ruledtabular}
\footnotetext[1]{Frozen core full configuration interaction calculation performed for Mg dimer.}
\end{table*}

\section{\label{sec:results-magnesium}MRSDCI Photoabsorption Spectra of Magnesium Clusters}

Next we present and discuss the results of our photoabsorption calculations for each isomer.

\subsection{Mg$_{2}$}

The simplest cluster of magnesium is Mg$_{2}$ with D$_{\infty h}$
point group symmetry. We obtained its CCSD optimized bond length to
be 3.93 \AA{} (\emph{cf.} Fig. \ref{fig:geometry-magnesium}\subref{subfig:subfig-mg2}),
which is in excellent agreement with the experimental value 3.89 \AA{} \cite{exp_mg_dimer_douglas}.
Using a DFT based methodology, several other theoretical values reported
are in excellent agreement with our optimized bond length of magnesium
dimer, i.e., Kumar and Car reported dimer bond length to be 3.88 \AA{}
\cite{car_kumar_magnesium_prb} using density functional molecular
dynamics with simulated annealing, Janecek \emph{et al.} computed
bond length to be 3.70 \AA{} \cite{wahl_mg_epjd} using DFT with LDA
approximation, 3.8 \AA{} bond length was reported by Stevens and Krauss
using multiconfiguration self-consistent field approach \cite{ground_excited_mg2_jcp},
3.91 \AA{} bond length of dimer was computed by Jellinek and Acioli
using DFT with BP86 exchange-correlation functional \cite{jellinek_mg2-mg5_jpca}, 
and Lyalin \emph{et al.} reported 3.926 \AA{} bond length using DFT
with B3LYP exchange-correlation functional \cite{ele-struct-magnesium-pra}.

\begin{figure}
\includegraphics[width=1\columnwidth]{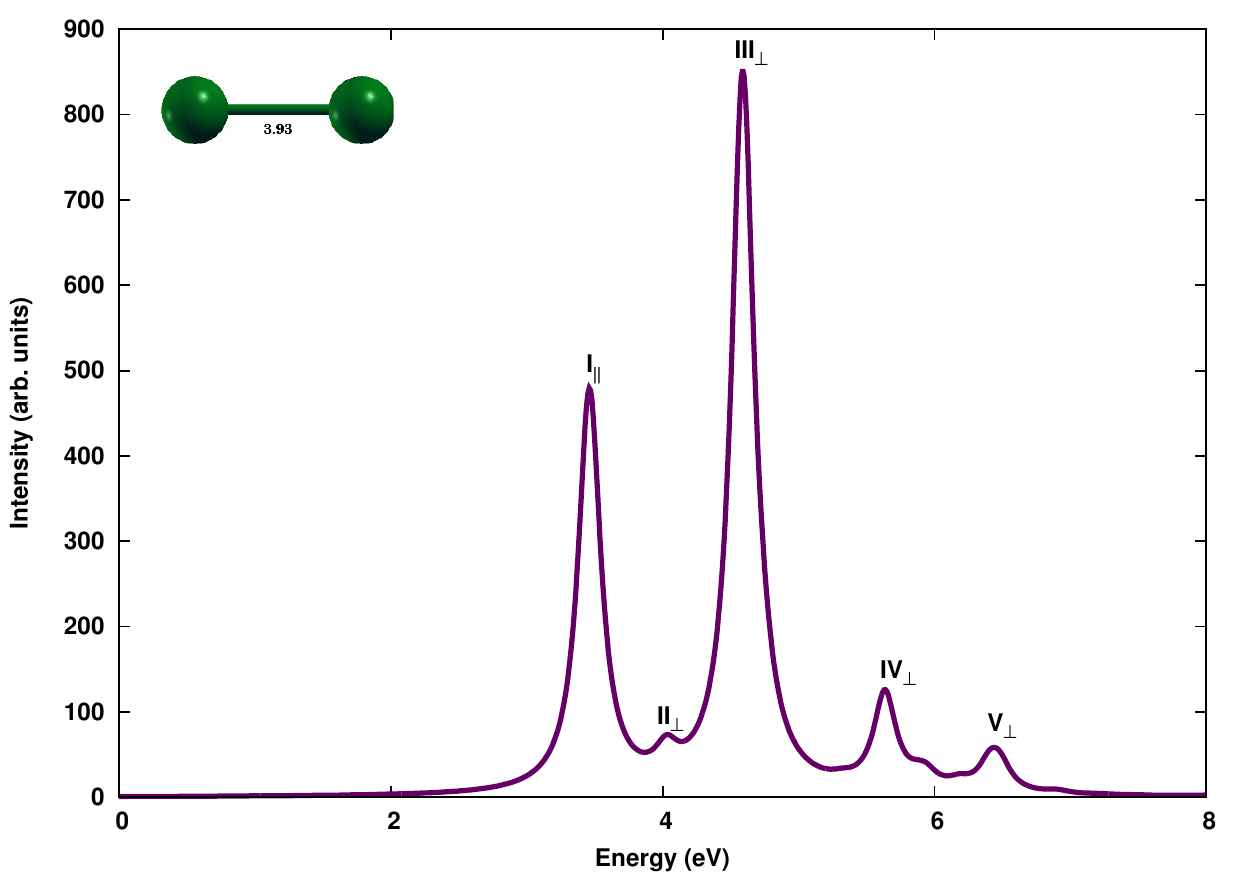}
\caption{\label{fig:plot-mg2-linear}The linear optical absorption spectrum of Mg$_{2}$, calculated using the MRSDCI approach.  The subscript $\parallel$ denotes the peak corresponding to the light polarization along the molecular axis, while the subscript $\perp$ labels those polarized perpendiculars to it. For plotting the spectrum, a uniform linewidth of 0.1 eV was used.}
\end{figure}

The computed photoabsorption spectra of Mg$_{2}$, as shown in Fig.
\ref{fig:plot-mg2-linear}, is characterized by a couple of intense
peaks in the 3 -- 5 eV range, and by weaker absorptions, in between,
and at higher energies.  Table I of Supplementary Material presents the many-body wave functions of excited
states contributing to various peaks  \cite{supplementary-material}.  The first peak at 3.46
eV, due to the absorption of longitudinally polarized photons, is
because of an excited state whose wave function is dominated by singly
excited configuration $H\rightarrow L+1$, where symbols H and L denote HOMO and LUMO orbitals respectively. This peak is reported in
the experimental photoabsorption at around 3.36 eV \cite{exp_mg_dimer_spectra_jcp,exp_mg_dimer_douglas}.
It is followed by a transversely polarized weaker absorption at 4.02
eV, characterized by several singly excited configurations, including
$H\rightarrow L+8$. The most intense peak occurs at 4.59 eV, whose
wave function is also dominated by single excitations such as $H-1\rightarrow L$
and $H\rightarrow L+3$. The location of this peak is in excellent
agreement with the experimental values of 4.59 eV reported by Lauterwald
 and Rademann \cite{mgdimerexp2,exp_mg_dimer_douglas}, and 4.62 eV measured by McCaffrey
and Ozin \cite{exp_mg_dimer_spectra_jcp}. 

The spectrum calculated using TDDFT by Solov'yov \emph{et al.}\cite{optical_mg_jpb} is in excellent agreement with our results. In their calculations,
the first peak is seen at 3.3 eV, followed by the most intense peak at 4.6 eV. The overall photoabsorption profile is also in accordance with our results.

\subsection{Mg$_{3}$}

We have optimized four low-lying geometries of magnesium trimer. The lowest energy structure at CCSD optimized level has equilateral triangular
shape with D$_{3h}$ symmetry and bond lengths of 3.48 \AA{}.   This agrees well with  other theoretical results reporting bond lengths, 3.51
\AA{} \cite{wahl_mg_epjd}, 3.48 \AA{} \cite{jellinek_mg2-mg5_jpca}, and 3.475 \AA{} \cite{ele-struct-magnesium-pra}. The next low-lying
isomer of magnesium trimer has a linear structure, with D$_{\infty h}$ symmetry. The optimized bond length is found to be 2.92 \AA{}. The
remaining two low-lying isomers have isosceles triangular shape, with C$_{2v}$ point group symmetry. Not much has yet been reported on
the bond lengths and electronic structure of these isomers.

\begin{figure}
\includegraphics[width=1\columnwidth]{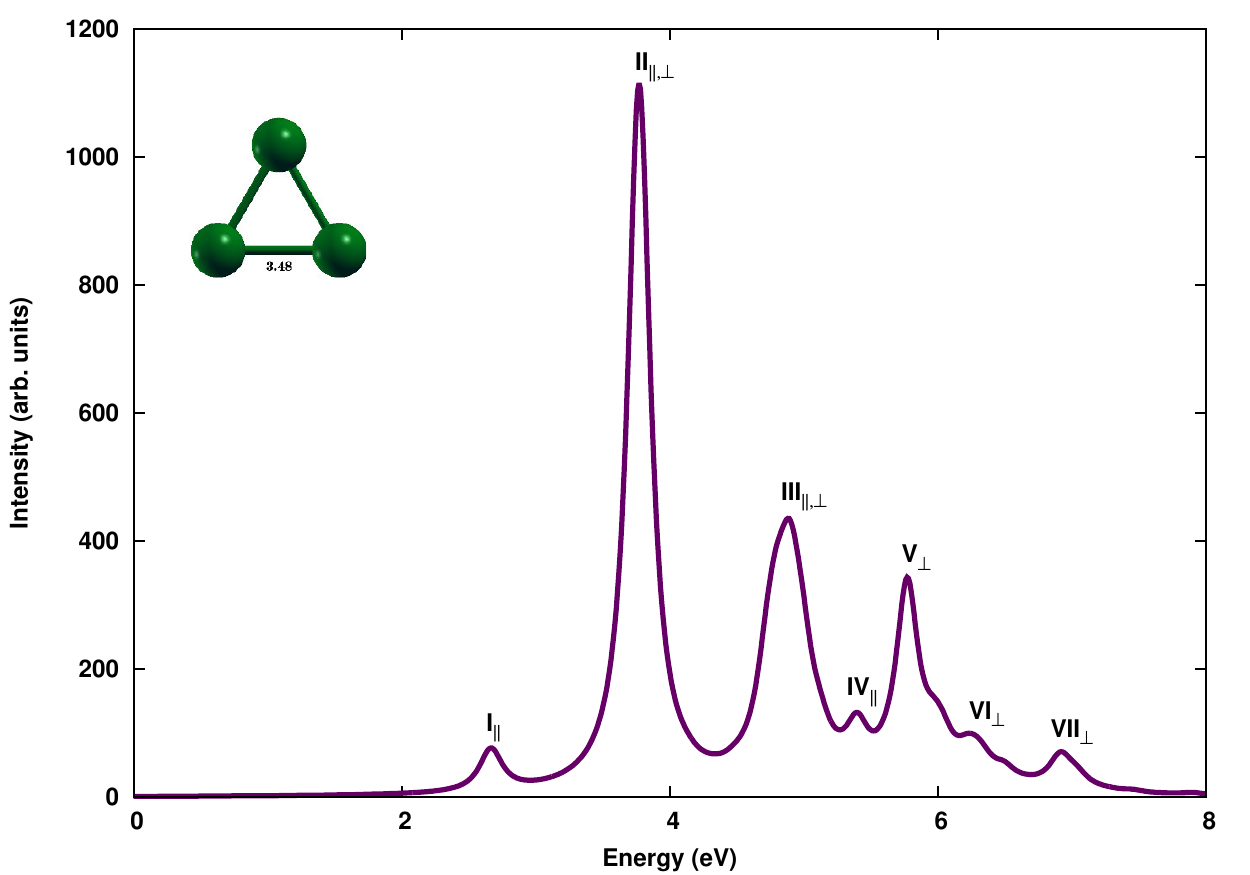}
\caption{\label{fig:plot-mg3-equil} The linear optical absorption spectrum of Mg$_{3}$ equilateral triangle isomer, calculated using the MRSDCI
approach. The peaks corresponding to the light polarized in the molecular plane are labeled with the subscript $\parallel$, while the subscript $\perp$ denotes
 those polarized perpendiculars to it. For plotting the spectrum, a uniform linewidth of 0.1 eV was used.}
\end{figure}

\begin{figure}
\includegraphics[width=1\columnwidth]{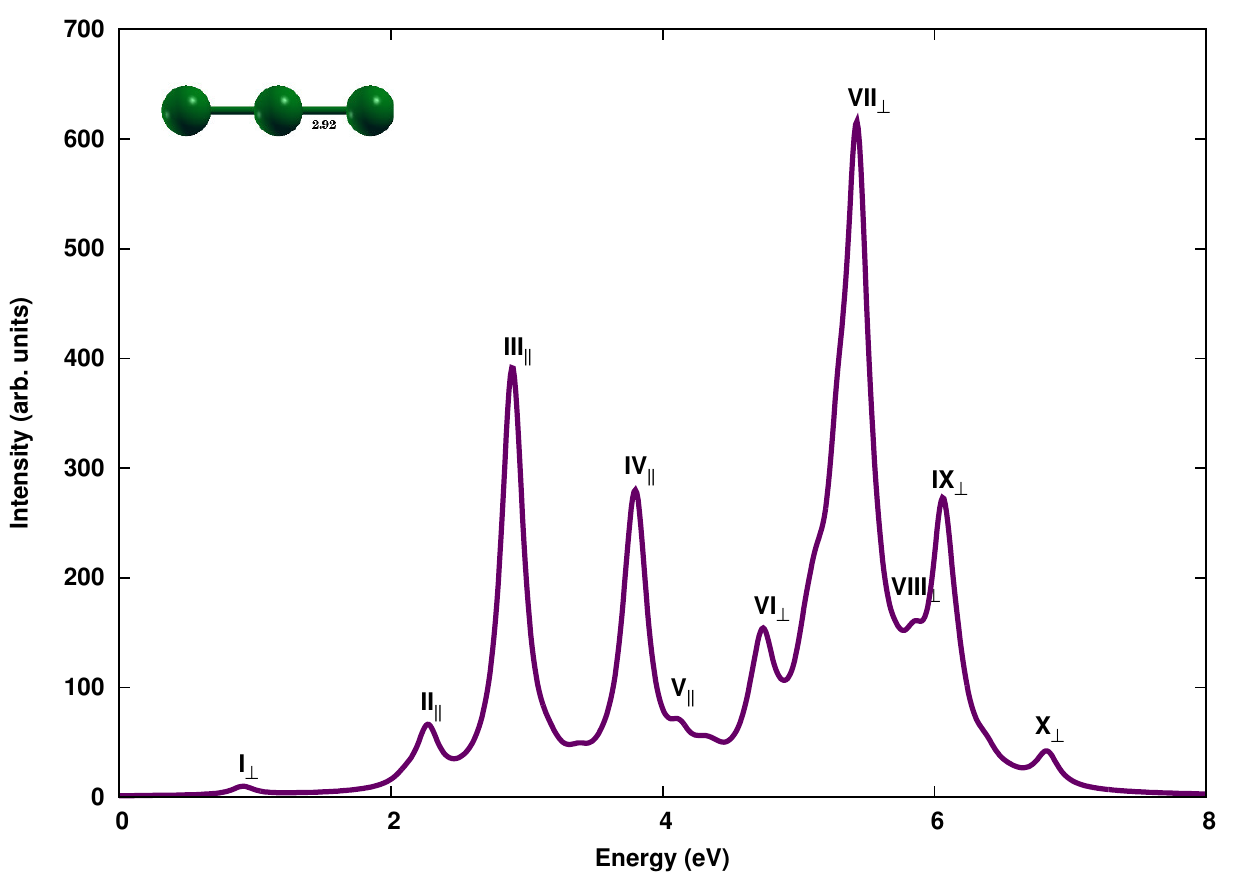}
\caption{\label{fig:plot-mg3-lin} The linear optical absorption spectrum of Mg$_{3}$ linear isomer, calculated using the MRSDCI approach. The peaks corresponding to the light polarized along the molecular axis are labeled with the subscript $\parallel$, while the subscript $\perp$ denotes those polarized perpendiculars to it. For plotting the spectrum, a uniform linewidth of 0.1 eV was used.}
\end{figure}

\begin{figure}
\includegraphics[width=1\columnwidth]{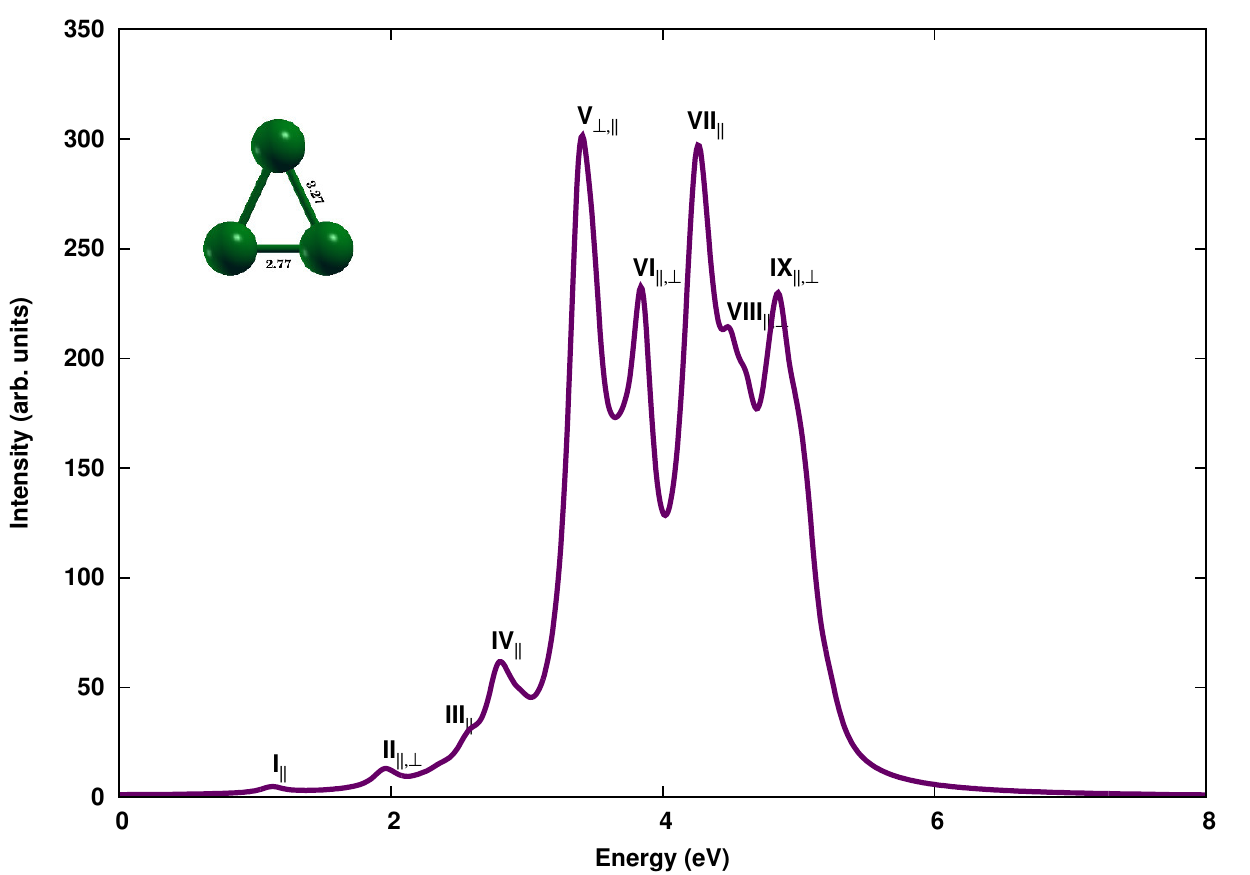}
\caption{\label{fig:plot-mg3-iso1} The linear optical absorption spectrum of Mg$_{3}$ isosceles triangle isomer-I, calculated using the MRSDCI approach. The peaks corresponding to the light polarized in the molecular plane are labeled with the subscript $\parallel$, while the subscript $\perp$ denotes those polarized perpendiculars to it. For plotting the spectrum, a uniform linewidth of 0.1 eV was used.  }
\end{figure}

Figs. \ref{fig:plot-mg3-equil}, \ref{fig:plot-mg3-lin}, \ref{fig:plot-mg3-iso1}, and \ref{fig:plot-mg3-iso2}, present the photoabsorption spectra of these isomers, while
Tables 2 -- 5 of  Supplementary Material \cite{supplementary-material} contain many-body wave functions of important excited states contributing  to various peaks. 

In the equilateral triangular isomer, the bulk of the oscillator strength is carried by a peak close to 3.75 eV.  The linear isomer shows an 
altogether different absorption spectrum with a number of peaks spread out in a wide energy range, with light polarized both parallel, and perpendicular, to the axis of the trimer. On the contrary, most of the oscillator strength in the absorption spectrum of isosceles triangular isomer-I is concentrated in the range of 3 -- 5 eV. The
lower-energy part of the spectrum of isosceles triangular isomer-II is somewhat red-shifted with respect to the isosceles isomer-I, while,
in the higher energy region, peaks are observed in the ultraviolet range.

The equilateral triangular isomer exhibits a weaker absorption peak at 2.6 eV, characterized by $H\rightarrow L$ and $H\rightarrow L+4$. This is followed by the most intense peak at 3.7 eV due to the light polarized both parallel and perpendicular to the plane of the isomer and with a dominant contribution from excitations
$H\rightarrow L$, $H\rightarrow L+2$, and $H-1\rightarrow L$. This is confirmed by an experimental measurement of photoabsorption of
Mg trimer in the argon matrix, which exhibits a peak at 3.64 eV \cite{exp_mg_dimer_spectra_jcp}. Semi-major peaks at around 4.7 eV and 5.8 eV obtain dominant contribution from single excitations $H\rightarrow L+7$, $H\rightarrow L+5$, and $H\rightarrow L+9$.  The latter peak is due to photons polarized perpendicular to the plane of isomer.

Comparing our results for the equilateral triangular isomer with the spectrum obtained by TDDFT calculations \cite{optical_mg_jpb}, we see very good agreement on the overall profile of spectrum and excitation energies. The first peak is observed at 2.5 eV, followed by the most intense one at 3.7 eV, in the TDDFT spectrum.\cite{optical_mg_jpb} Excitation energies and relative oscillator strengths are also in good agreement with our results.

Because the ground state of Mg$_{3}$ linear isomer is a spin triplet, its many-particle wave function predominantly consists of a configuration
with two degenerate singly occupied molecular orbitals, referred to as $H_{1}$ and $H_{2}$ in rest of the discussion. The linear trimer
of magnesium cluster exhibits absorption in the entire energy range explored. Very feeble peaks are observed at 0.9 eV and 2.3 eV, due
to states dominated by single excitations $H_{1}\rightarrow L+8$, $H_{1}\rightarrow L+2$, and $H_{1}\rightarrow L+4$. The wave function
of the state leading to the second most intense peak at 2.9 eV is dominated by the configuration $H_{1}\rightarrow L+3$. The state
leading to the most intense peak at 5.4 eV derives almost equal contributions from configurations $H-2\rightarrow L$ and $H-1\rightarrow L+2$.
The absorption due to the longitudinally polarized light contributes to the lower energy part of the spectrum, while transversely polarized
light contributes to the remaining higher energy part of the spectrum.

Both isosceles triangular isomers have a spin triplet ground state; hence their excited state wave functions will consist of configurations
involving electronic excitations from singly occupied degenerate $H_{1}$ and $H_{2}$ molecular orbitals, in addition to other doubly occupied
orbitals. In the case of isosceles triangular isomer - I (\emph{cf}. Fig. \ref{fig:plot-mg3-iso1}), the spectrum starts with a very feeble
peak at 1.13 eV, leading to a state whose wave function derives the main contribution from $H_{1}\rightarrow L+1$ configuration. However,
most of the absorption takes place in the energy range of 3 -- 5 eV, with two equally intense peaks at 3.4 eV (peak V) and 4.2 (peak VII)
eV, while the other two peaks (VI and IX) and a shoulder (VIII) in that range, are also quite intense. Peak V is due to three closely-spaced
states, the first of which is reached by photons polarized perpendicular to the plane of the triangle, while the other two are due to photons polarized
in the plane of the cluster. Wave functions of all the three states derive dominant contributions from singly-excited configurations. The other
most intense peak (VII) is due to two closely located states and displays mixed polarization characteristics. Wave functions of both
these excited states, in addition to the single excitations, derive important contributions from doubly excited configurations as well.

The absorption spectrum of the isosceles triangular isomer -II (\emph{cf}. Fig. \ref{fig:plot-mg3-iso2}) appears red-shifted as compared to
that of the previous isomer and exhibits a set of well-separated peaks. There is just one excited state contributing to the most intense
peak at 2.6 eV (peak V), which is due to the absorption of a photon polarized in the plane of the triangle. The wave function of this state
mostly consists of the configuration $H_{2}\rightarrow L$, with some contribution from a doubly-excited configuration $H_{2}\rightarrow L;H_{1}\rightarrow L+10$. Two almost equally intense peaks of absorption due to in-plane polarized photons occur at 3.5 eV (peak VII) and 3.9 eV (peak VIII). One excited
state each contributes to these peaks, and wave functions of these states are dominated by singly-excited configurations which include
$H-2\rightarrow H_{1}$, $H_{1}\rightarrow L+7$, and $H_{1}\rightarrow L+17$. This isomer also exhibits a strong mixing of doubly excited configurations
for excited states contributing to higher energy peaks. Significant differences in the optical absorption spectra of the two isosceles
triangle shaped isomers point to a strong structure-property relationship when it comes to optical properties of these clusters.

\begin{figure}
\includegraphics[width=1\columnwidth]{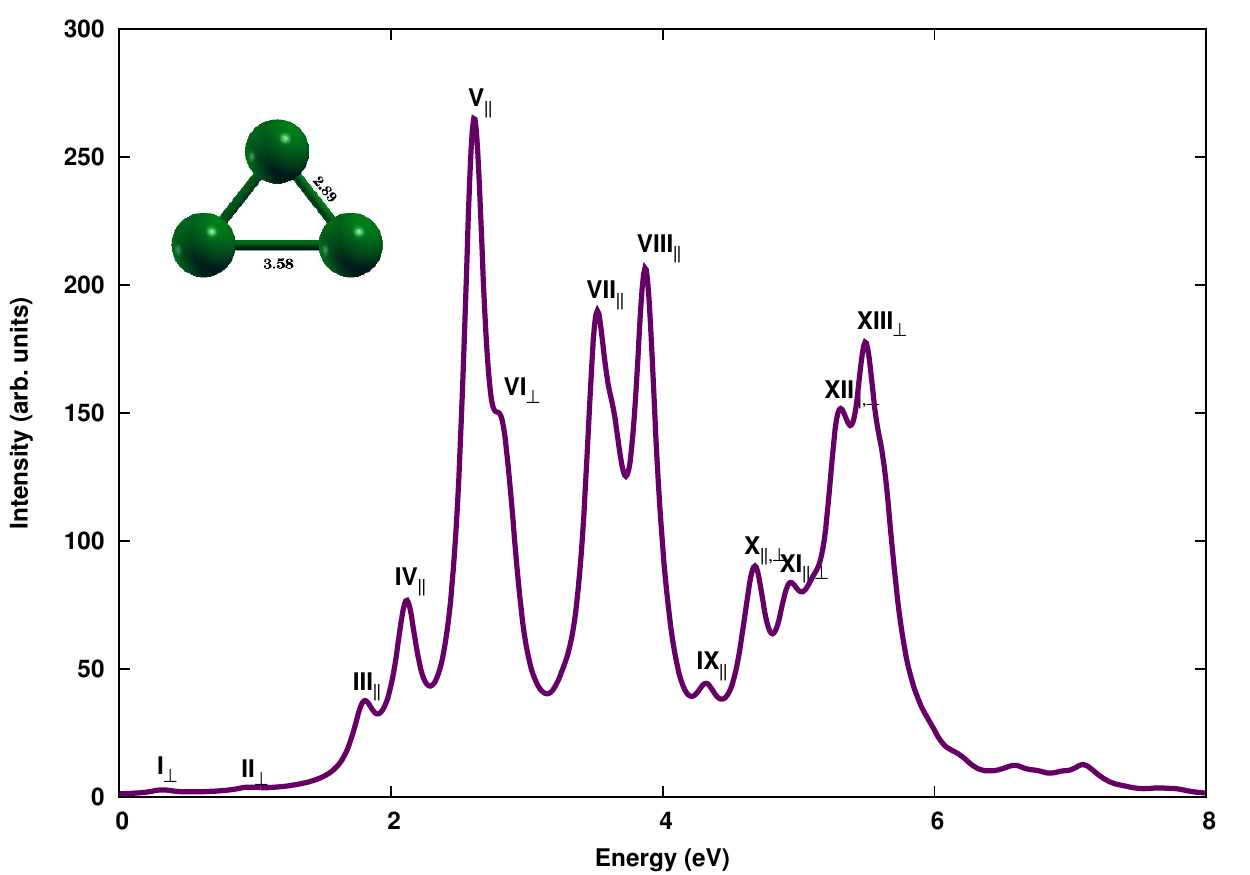}
\caption{\label{fig:plot-mg3-iso2}The linear optical absorption spectrum of Mg$_{3}$ isosceles triangle isomer-II, calculated using the MRSDCI approach. The peaks corresponding to the light polarized in the molecular plane are labeled with the subscript $\parallel$, while the subscript $\perp$ denotes those polarized perpendiculars to it. For plotting the spectrum, a uniform linewidth of 0.1 eV was used.}
\end{figure}


\subsection{Mg$_{4}$}

The most stable isomer of Mg$_{4}$ cluster has a closed-shell electronic ground state, with the structure of a perfect tetrahedron (\emph{cf.}
Fig. \ref{fig:geometry-magnesium}\subref{subfig:subfig-mg4-pyramidal}), corresponding to T$_{d}$ point group symmetry, which, henceforth, we refer to as a pyramid. We computed
the optimized bond length to be 3.22 \AA{}, which agrees well with the previously reported values for this structure 3.09 \AA{} \cite{ahlrichs_mg_pccp},
3.33 \AA{} \cite{wahl_mg_epjd}, 3.18 \AA{} \cite{jellinek_mg2-mg5_jpca}, 3.31 \AA{} \cite{manninen_mg_evolution_epjd}, and 3.32 \AA{} \cite{kaplan_mg3_jcp}.
The rhombus isomer (\emph{cf.} Fig. \ref{fig:geometry-magnesium}\subref{subfig:subfig-mg4-rhombus}) with D$_{2h}$ point group symmetry, and bond length of 3.0 \AA{}, along with the acute angle 63.5\degree, has $^{3}B_{3u}$ electronic ground state, which is 1.02 eV higher than the global minimum structure.
Square isomer (\emph{cf.} Fig. \ref{fig:geometry-magnesium}\subref{subfig:subfig-mg4-square}) with D$_{4h}$ point group symmetry, and an optimized bond length of 3.06
\AA{}, has $^{3}A_{g}$ electronic ground state, which is energetically 1.37 eV higher than the most stable structure.

\begin{figure}
\includegraphics[width=1\columnwidth]{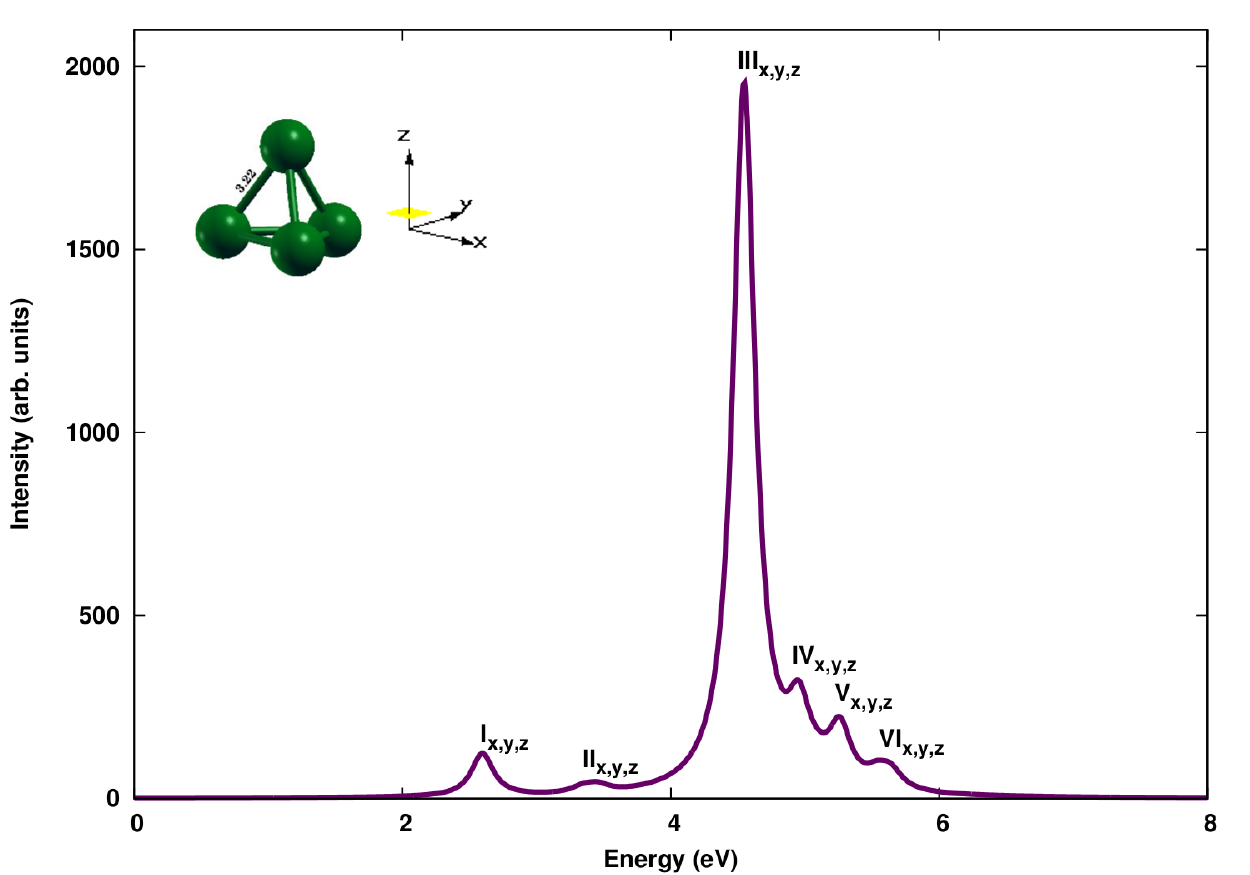}
\caption{\label{fig:plot-mg4-pyra} The linear optical absorption spectrum of pyramidal Mg$_{4}$ isomer, calculated using the MRSDCI approach.
The peaks corresponding to the light polarized along the Cartesian axes are labeled accordingly. For plotting the spectrum, a uniform
linewidth of 0.1 eV was used.} 
\end{figure}

\begin{figure}
\includegraphics[width=1\columnwidth]{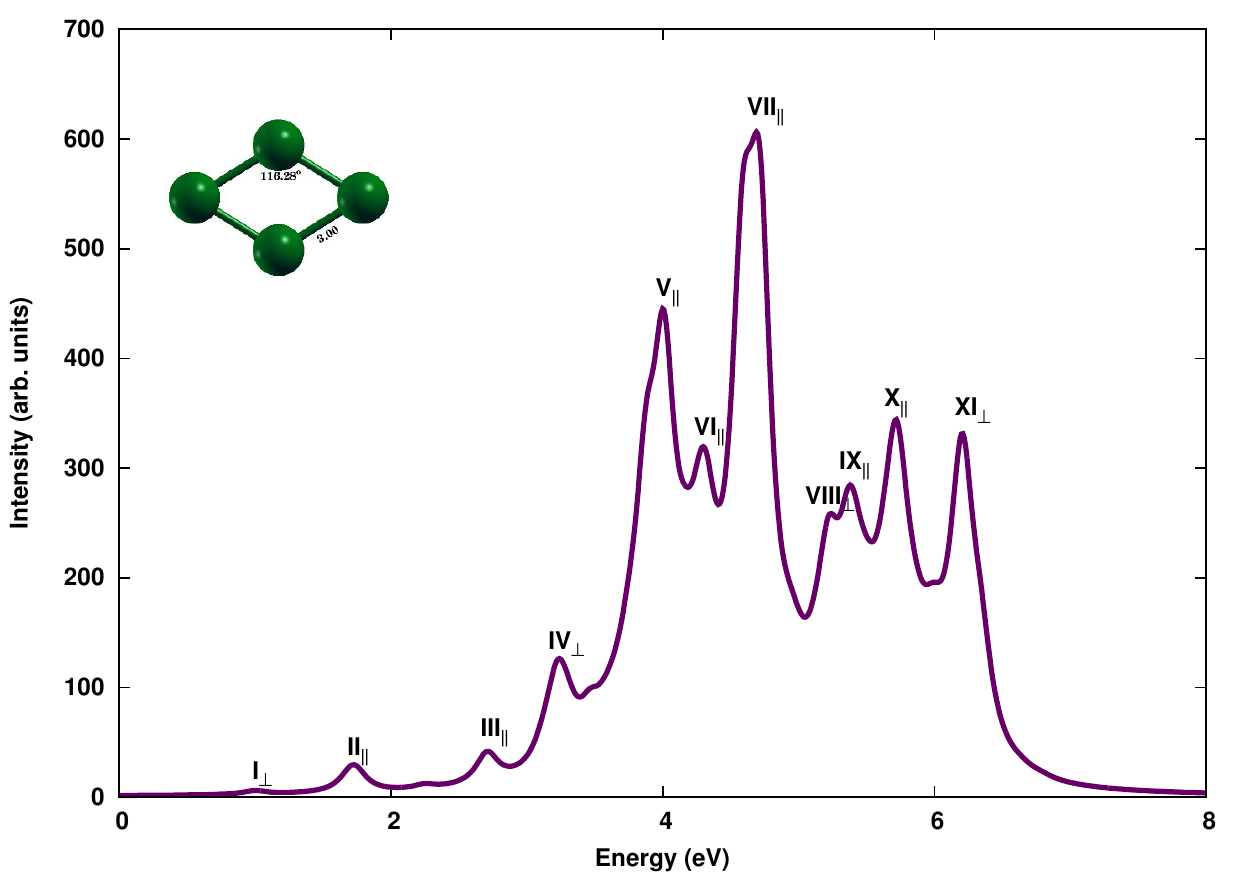}
\caption{\label{fig:plot-mg4-rho} The linear optical absorption spectrum of rhombus Mg$_{4}$, calculated using the MRSDCI approach. The peaks corresponding to the light polarized in the molecular plane are labeled with the subscript $\parallel$, while the subscript $\perp$ denotes those polarized perpendiculars to it. For plotting the spectrum, a uniform linewidth of 0.1 eV was used.}
\end{figure}

\begin{figure}
\includegraphics[width=1\columnwidth]{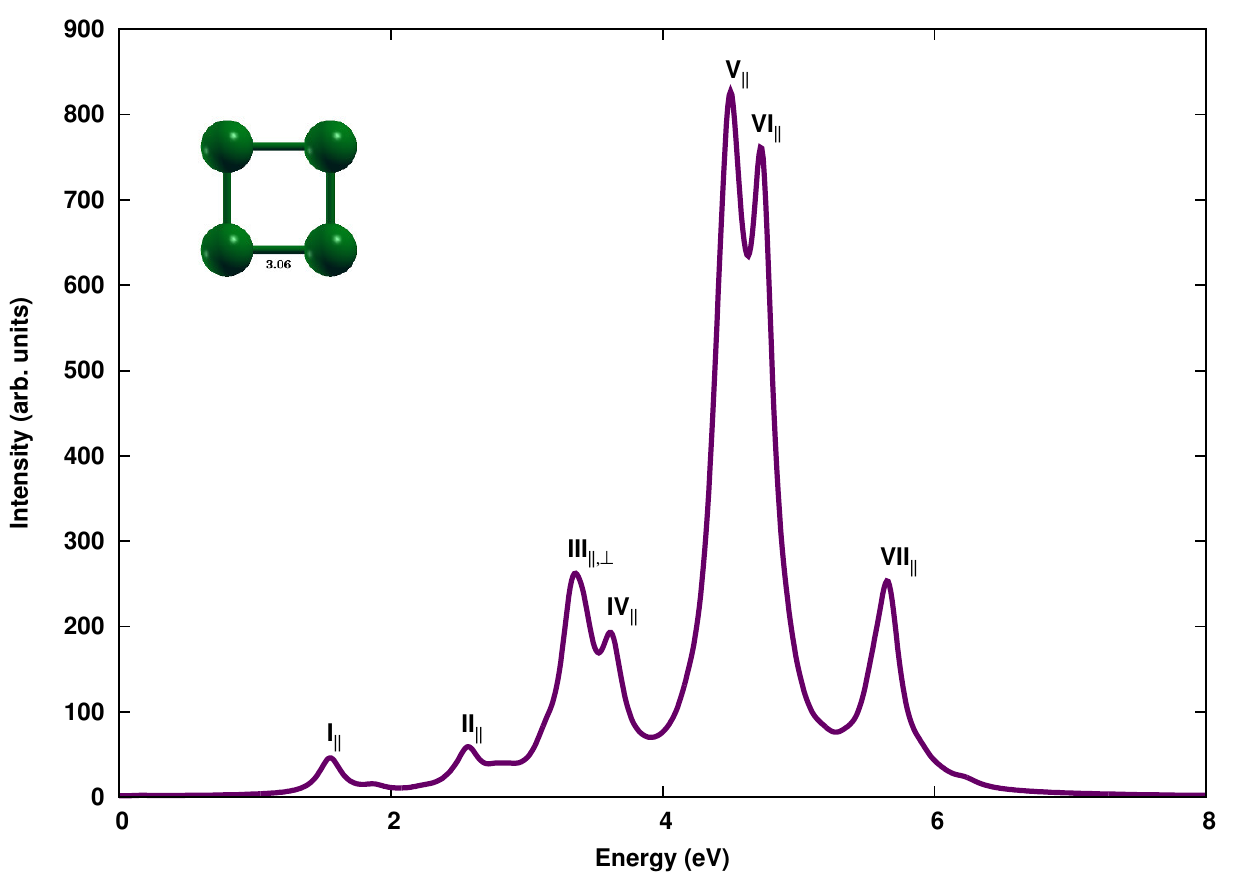}
\caption{\label{fig:plot-mg4-sqr} The linear optical absorption spectrum of square Mg$_{4}$, calculated using the MRSDCI approach. The peaks corresponding to the light polarized in the molecular plane are labeled with the subscript $\parallel$, while the subscript $\perp$ denotes those polarized perpendiculars to it. For plotting the spectrum, a uniform linewidth of 0.1 eV was used.}
\end{figure}

The absorption spectra of pyramidal, rhombus and square isomers are presented in Figs. \ref{fig:plot-mg4-pyra}, \ref{fig:plot-mg4-rho},
and \ref{fig:plot-mg4-sqr}, respectively, while the many-particle wave functions of the excited states contributing to various peaks
are presented in Tables 6, 7 and 8, respectively, of the Supplemental Material \cite{supplementary-material}.

Because of the three-dimensional structure of the pyramidal isomer, all three Cartesian components contribute to the transition dipole
moments, thereby implying a three-dimensional polarization of the incident photons with respect to the chosen coordinate system. The
onset of absorption in this isomer occurs at 2.6 eV, due to a state whose many-particle wave function is dominated by configurations $H-1\rightarrow L$,
$H\rightarrow L$, and $H-2\rightarrow L$. The most intense peak in the spectrum is located at 4.54 eV due to a state whose wave function
is dominated by several single excitations such as $H\rightarrow L+2$, $H-1\rightarrow L+1$, and $H-1\rightarrow L+3$\emph{ }etc. The TDDFT
absorption spectrum of this isomer reported by Solov'yov \emph{et al} \cite{optical_mg_jpb}. is slightly red-shifted compared our calculated
spectrum, however, its absorption pattern is similar to ours in that a single most intense peak at 4.2 eV is followed by several less intense
peaks at higher energies.

In the case of rhombus-shaped isomer, optical absorption starts with a weak peak at a rather low energy close to 1.00 eV, with the bulk of
the oscillator strength distributed in the energy range 4 -- 6 eV, consisting of several equally intense and close-lying peaks. The first
weak peak at 0.98 eV corresponds to a photon polarized perpendicular to the plane of the molecule and is due to a state dominated by configuration
$H_{1}\rightarrow L+1$. The most intense peak at 4.7 eV is due to a photon polarized in the plane of the isomer, and the corresponding
excited state wave function is dominated by configurations $H-1\rightarrow L+1$ and $H-2\rightarrow L$. It is preceded by a shoulder at 4.6 eV, with
identical polarization properties, and an excited state wave function which derives contributions from configurations $H_{2}\rightarrow L+8$
and $H-1\rightarrow L$. The most intense peak corresponding to perpendicular polarization is located at 6.22 eV, with the excited state wave function
dominated by doubly-excited configurations.

The inversion symmetry of the ground state of the square isomer is just opposite to that of the rhombus structure (\emph{cf.} Fig. \ref{fig:geometry-magnesium}),
so that, as per dipole selection rule, the excited states contributing to the linear absorption spectra for the two structures also have
opposite inversion symmetries. Quantitatively speaking, the absorption spectrum of the square structure is slightly blue-shifted as compared
to the rhombus, and red-shifted as compared to pyramidal isomer, with the majority of absorption occurring in the energy range 3--6 eV.
The onset of absorption spectrum occurs at 1.55 eV with a peak due to the light polarized in the plane of isomer, leading to a state
whose wave function is a mixture predominantly of configurations $H_{1}\rightarrow L+2$, $H_{1}\rightarrow L+10$, and $H_{1}\rightarrow L+15$. This isomer,
similar to the case of the rhombus, exhibits two very closely spaced high-intensity peaks, located at 4.50 eV and 4.73 eV, both of which are due to the
absorption of photons polarized in the plane of the cluster. The first of these peaks (peak V) is due to a state whose wave function is a
mixture with almost equal contributions from single excitations as $H-1\rightarrow L$, $H_{1}\rightarrow L+2$, and $H_{1}\rightarrow L+15$,
and also a doubly-excited configuration. The excited state causing the second one (peak VI) is dominated by singly excited configurations
$H_{1}\rightarrow L+20$ and $H_{1}\rightarrow L+24$. The last peak of the computed spectrum (peak VII) has a relatively lower intensity and is due to a state dominated by single excitations $H-2\rightarrow L+13$ and $H_{2}\rightarrow L+20$. 

\subsection{Mg$_{5}$}

We optimized geometries of two isomers of Mg$_{5}$: (a) bipyramid with the D$_{3h}$ symmetry and (b) a pyramidal structure with the
C$_{4v}$ point group symmetry. The lowest lying bipyramidal isomer has $^{1}$A$_{1}^{'}$ electronic ground state and is just 0.18
eV lower in energy as compared to the pyramid structure. Our optimized geometry for the bipyramid has bond lengths of 3.15 \AA{} and
3.52 \AA{}, as against 3.00 \AA{}, 3.33 \AA{} reported by J. Jellinek and Acioli \cite{jellinek_mg2-mg5_jpca}, and 3.09 \AA{}, 3.44 \AA{}
reported by Andrey \emph{et al} \cite{ele-struct-magnesium-pra}.

The bipyramidal isomer of Mg$_{5}$ cluster exhibits an absorption spectrum very different from other isomers, as displayed in Fig. \ref{fig:plot-mg5-biprism}, while Table 9 of Supplemental Material \cite{supplementary-material} presents the many-particle wave functions of the excited states contributing to various peaks. The optical absorption spectrum of bipyramid Mg$_{5}$ has no absorption until 3.5 eV, while most of the absorption takes place in a narrow
energy range 5.3 -- 6.3 eV. The absorption spectrum begins at 3.6 eV through a photon polarized in the basal plane of the bipyramid,
with a very feeble peak corresponding to a state whose wave function derives the main contribution from $H-1\rightarrow L+4$ configuration.
This is followed by several such smaller peaks. The most intense peak at 5.4 eV has a dominant contribution from $H\rightarrow L+1$ along
with other singly excited configurations, with absorption polarized again along the basal plane of the pyramid. A shoulder at 5.6 eV,
however, corresponds to the absorption of light with polarization along the $z-$ direction, which is perpendicular to the basal plane. This feature is caused
by an excited state whose wave function is mainly a linear combination of several singly-excited configurations. The TDDFT spectrum computed
by Solov'yov \emph{et al} \cite{optical_mg_jpb}. shows optical activity in the energy range of 2--4 eV, which is not observed in our calculated
spectrum. However, a quasi-continuous spectrum is seen at higher energies in both calculations.

\begin{figure}
\includegraphics[width=1\columnwidth]{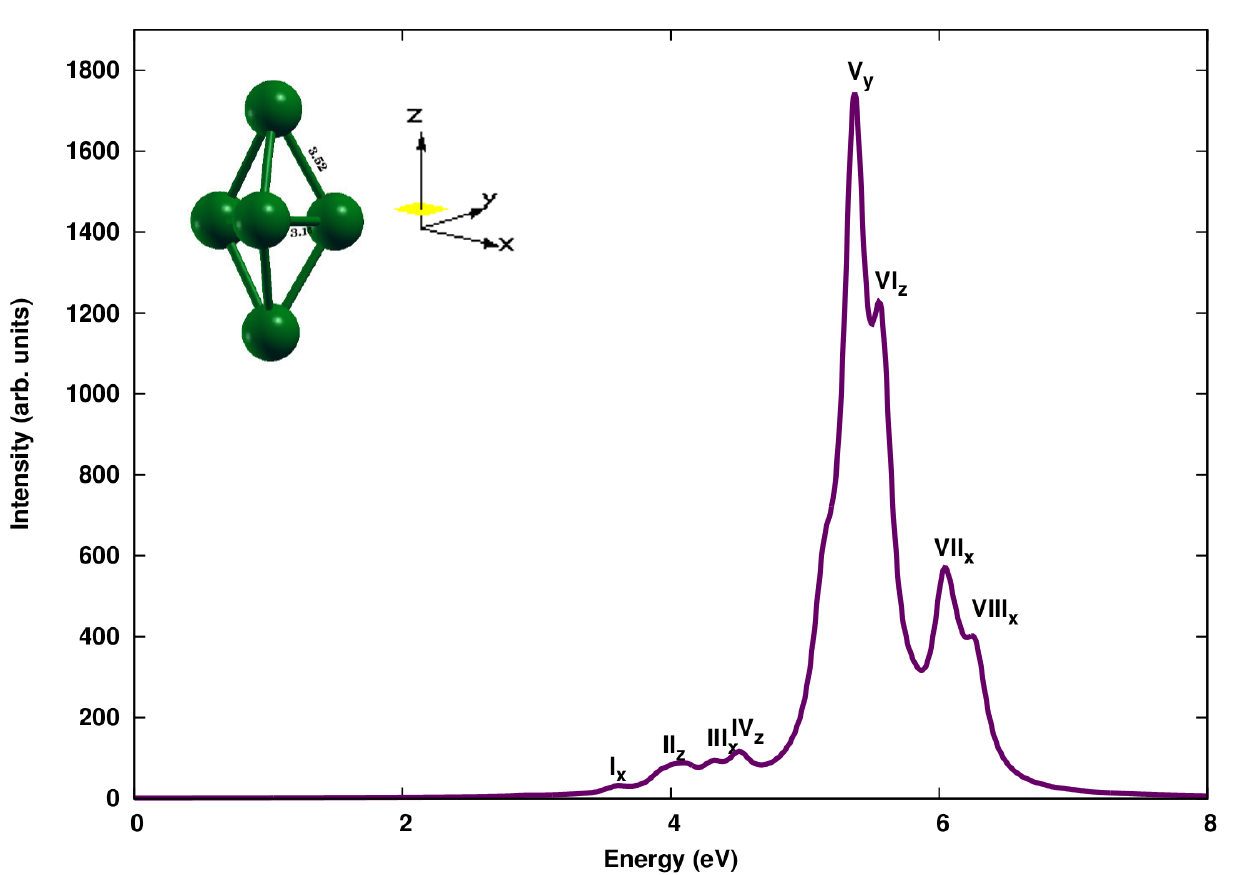}
\caption{\label{fig:plot-mg5-biprism} The linear optical absorption spectrum of bipyramidal Mg$_{5}$ isomer, calculated using the MRSDCI approach.
The peaks corresponding to the light polarized along the Cartesian axes are labeled accordingly. For plotting the spectrum, a uniform
linewidth of 0.1 eV was used.}
\end{figure}

The optical absorption spectrum of pyramid shaped isomer, to the best of our knowledge, has not been computed, so far, by any other author.
The entire absorption spectrum of the pyramid shaped isomer is highly red-shifted as compared to the bipyramid isomer. The many-particle
wave functions of excited states contributing to the peaks are presented in Table 10 of Supplemental Material \cite{supplementary-material}.
The optical absorption of this isomer exhibits a few feeble peaks in the low energy range,
with the onset of spectrum occurring at 2.24 eV through photons polarized in the basal plane of the pyramid, as well as perpendicular to it. The wave functions of the two excited states contributing to this peak are dominated by configurations $H-1\rightarrow L$ and
$H\rightarrow L+2$. The second most intense peak close to 3.5 eV is well separated from the most intense one located at 4.2 eV. The
former has mixed polarization characteristics, with two states dominated by configurations $H-1\rightarrow L+3$ and $H-2\rightarrow L$, besides
several other single excitations. The most intense peak is due to light polarized perpendicular to the basal plane of the pyramid, and
wave function of the excited state involved is dominated by configurations $H-2\rightarrow L+1$ and $H-2\rightarrow L+3$. The last absorption
peak in the probed energy range  is located at 6.27 eV, caused by a photon polarized perpendicular to the base of the pyramid, and is
due to an excited state deriving the main contribution from a doubly-excited configuration, along with several single excitations. Pyramid shaped
isomer exhibits prominent optical absorption in the higher energy range, with almost regularly spaced peaks of declining intensities,
in contrast to single major peak observed in the spectrum of bipyramidal isomer. These differences can help in experimental identification of geometries
of various isomers, through optical absorption spectroscopy.

\begin{figure}
\includegraphics[width=1\columnwidth]{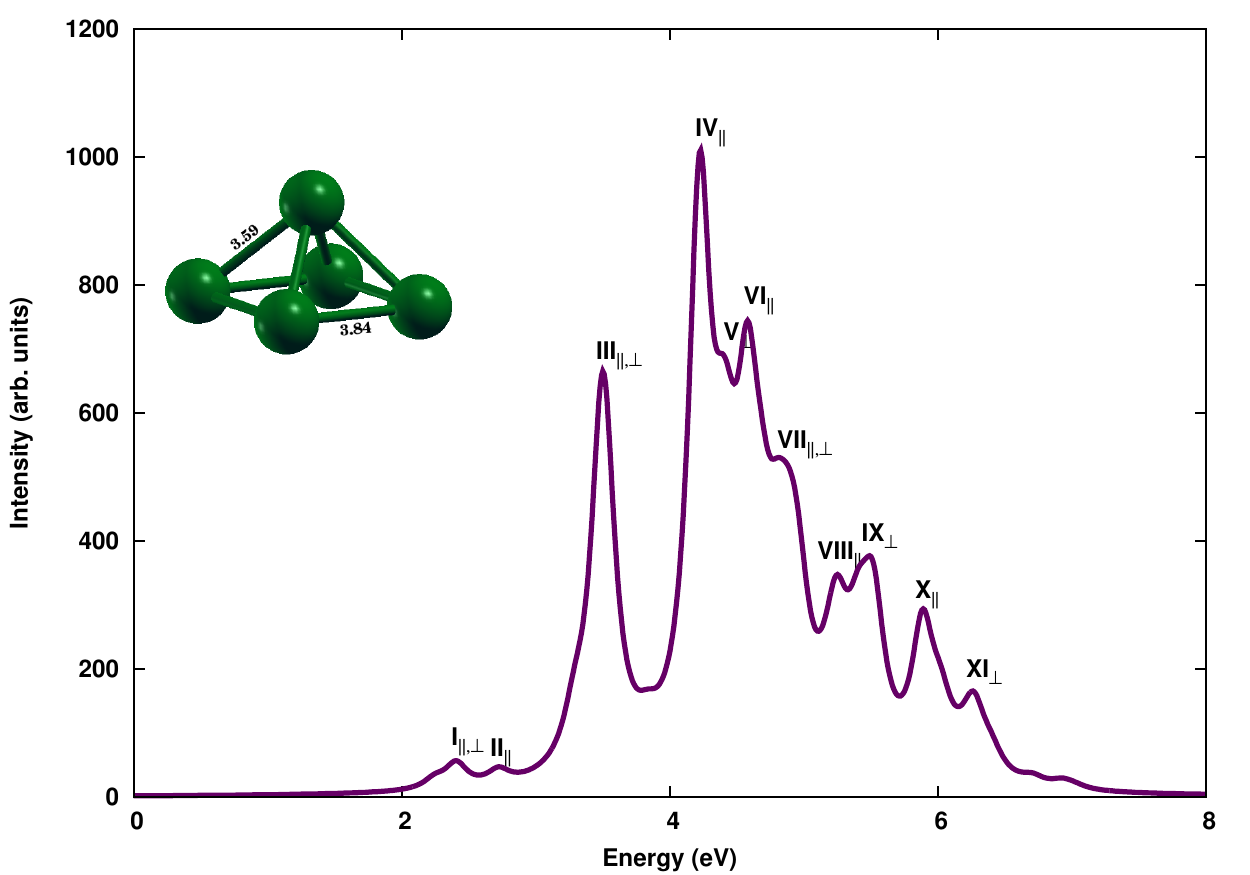}
\caption{\label{fig:plot-mg5-param} The linear optical absorption spectrum of pyramidal Mg$_{5}$, calculated using the MRSDCI approach. The peaks corresponding to the light polarized along the base of the pyramid are labeled with the subscript $\parallel$, while the subscript $\perp$ denotes those polarized perpendiculars to it. For plotting the spectrum, a uniform linewidth of 0.1 eV was used.}
\end{figure}

\section{\label{sec:conclusions}CONCLUSIONS AND OUTLOOK}

In this work, large-scale first-principles electron correlated calculations of photoabsorption spectra of several low-lying isomers of magnesium clusters Mg$_{n}$, (n=2--5), were presented. For the case of magnesium dimer, we employed one of the best possible electronic structure methods, namely FCI method, within the frozen-core approximation, to compute its electronic states. Calculations for the remaining clusters were performed using MRSDCI approach, which takes excellent account of the electron-correlation effects both for ground and excited states. We have also analyzed the nature of the many-particle wave functions of the excited states visible in the absorption spectra.  Distinct signature spectra are exhibited by isomers of a given cluster, suggesting a strong structure-property relationship. This behavior can be utilized in the experiments to distinguish between different isomers of a cluster, using optical absorption spectroscopy.  Given the fact electron-correlation effects were included in our calculations in a sophisticated manner by means of large-scale CI expansions, we believe that our results can be used as theoretical benchmarks of absorption spectra of Mg clusters, against which both the experimental and other theoretical results can be compared. We hope that our work will lead to experimental measurements of the optical absorption spectra of magnesium clusters of various shapes and sizes.

\acknowledgments
R.S. acknowledges the Council of Scientific and Industrial Research (CSIR) and Science and Engineering Research Board (SERB) India, for research fellowships (09/087/(0600)2010-EMR-I), (PDF/2015/000466). Authors kindly acknowledge computational resources provided by National Param Yuva Supercomputing Facility, C-DAC, Pune. 

\bibliography{mg_clus}

\end{document}